\documentclass[aps,prc,twocolumn,groupedaddress,showpacs,superscriptaddress,nofootinbib]{revtex4-2}
\usepackage{graphicx,amsmath}
\usepackage{amssymb}
\usepackage{amsmath}
\usepackage{tablefootnote}
\usepackage{units}

\usepackage[dvipsnames]{xcolor}

\usepackage{stix}

\usepackage{tikz}
\usepackage{threeparttable,lipsum}
\usepackage{mathtools}
\usepackage{hyperref}
\usepackage{dcolumn}
\usepackage{bm}

\begin{document}


\title{Thermonuclear reaction rate of $^{29}$Si(p,$\gamma$)$^{30}$P}


\author{L. N. Downen}
\affiliation{Department of Physics \& Astronomy, University of North Carolina at Chapel Hill, NC 27599-3255, USA}
\affiliation{Triangle Universities Nuclear Laboratory (TUNL), Duke University, Durham, North Carolina 27708, USA}

\author{C. Iliadis}
\email[]{iliadis@unc.edu}
\affiliation{Department of Physics \& Astronomy, University of North Carolina at Chapel Hill, NC 27599-3255, USA}
\affiliation{Triangle Universities Nuclear Laboratory (TUNL), Duke University, Durham, North Carolina 27708, USA}

\author{A. E. Champagne}
\affiliation{Department of Physics \& Astronomy, University of North Carolina at Chapel Hill, NC 27599-3255, USA}
\affiliation{Triangle Universities Nuclear Laboratory (TUNL), Duke University, Durham, North Carolina 27708, USA}

\author{T. B. Clegg}
\affiliation{Department of Physics \& Astronomy, University of North Carolina at Chapel Hill, NC 27599-3255, USA}
\affiliation{Triangle Universities Nuclear Laboratory (TUNL), Duke University, Durham, North Carolina 27708, USA}

\author{A. Coc}
\affiliation{CNRS/IN2P3, IJCLab, Universit\'e Paris-Saclay, B\^atiment, 104, F-91405 Orsay Campus, France}

\author{J. Dermigny}
\affiliation{Department of Physics \& Astronomy, University of North Carolina at Chapel Hill, NC 27599-3255, USA}
\affiliation{Triangle Universities Nuclear Laboratory (TUNL), Duke University, Durham, North Carolina 27708, USA}


\date{\today}

\begin{abstract}
The thermonuclear rate of the $^{29}$Si(p,$\gamma$)$^{30}$P reaction impacts the $^{29}$Si abundance in classical novae. A reliable reaction rate is essential for testing the nova paternity of presolar stardust grains. At present, the fact that no classical nova grains have been unambiguously identified in primitive meteorites among thousands of grains studied is puzzling, considering that classical novae are expected to be prolific producers of dust grains. We investigated the $^{29}$Si $+$ $p$ reaction at center-of-mass energies of $200$ $-$ $420$~keV, and present improved values for resonance energies, level excitation energies, resonance strengths, and branching ratios. One new resonance was found at a center-of-mass energy of $303$ keV. For an expected resonance at $215$~keV, an experimental upper limit could be determined for the strength. We evaluated the level structure near the proton threshold, and present new reaction rates based on all the available experimental information. Our new reaction rates have much reduced uncertainties compared to previous results at temperatures of $T$ $\ge$ $140$~MK, which are most important for classical nova nucleosynthesis. Future experiments to improve the reaction rates at lower temperatures are discussed.
\end{abstract}


\maketitle


\section{Introduction}\label{sec:intro}
Classical novae result from the accretion of hydrogen-rich material onto the surface of a white dwarf in a close binary system (see Ref.~\cite{Jose2016} for a review). The transferred matter does not fall directly onto the white dwarf, but forms an accretion disk around the compact star. Part of this matter moves inward and accumulates on top of the white dwarf, where it is gradually compressed and heated, until nuclear reactions begin to occur at the base of the accreted envelope. When the generated nuclear energy becomes too large to be transported by radiation, a thermonuclear runaway (TNR) ensues with the onset of convection. This is caused by a thin-shell instability, aided by the partial degenerate nature of the matter taking part in the nuclear burning. As a result, material is ejected into the interstellar medium at high velocities, giving rise to a classical nova.

While the basic mechanism has been known for a long time \cite{Schatzmann1949}, numerical models attempting to quantitatively reproduce classical nova observables face many obstacles related to both hydrodynamical effects and uncertain nuclear reaction rates. For example, how the TNR is initiated is far from clear. A related question pertains to the mixing of accreted, solar-like matter with outer layers of the white dwarf. While metallicity enhancements observed in nova ejecta point to significant mixing at the core-envelope interface during the TNR, the nature of the mechanism is not understood \cite{jose2020}.

Classical novae are observed at all wavelengths, ranging from radio waves to $\gamma$ rays, depending on the time since the TNR \cite{chomiuk2021}. The observations are important for understanding the energetics, mass loss, and shocks associated with these events. Spectroscopically inferred elemental abundances in nova ejecta, which are impacted by prior nuclear burning, also provide valuable information, although the abundance estimates are subject to significant uncertainties \cite{jose2008,downen2012}. Another intriguing potential probe of nova nucleosynthesis comes from presolar stardust gains that are embedded in primitive meteorites \cite{nittler2016}. Many classical novae are prolific producers of both carbon-rich and oxygen-rich dust \cite{gehrz1998} and, therefore, the isotopic composition of several elements in the dust grains will reflect the hydrodynamical conditions and mixing processes during nuclear burning \cite{starrfield2008}. However, while several authors have suggested a nova paternity for specific stardust grains, no grains from novae have been unambiguously identified yet. Therefore, they are referred to in the literature as {\it ``nova candidate grains.''}

Such grains are mainly composed of either SiC, silicate, graphite, or oxide (see Tab.~2 of Ref.~\cite{iliadis2018}). For the first three groups, measured silicon isotopic ratios are available, in addition to isotopic ratios for carbon, nitrogen, oxygen, or sulfur. The data provide important information for associating a particular grain with a classical nova paternity and for constraining the mechanisms of the explosion. The measured silicon isotopic abundances, however, can only serve as a useful probe if the thermonuclear reaction rates are known reliably. 

Recently, Lotay {\it et al.} \cite{Lotay2020} estimated the $^{29}$Si $+$ $p$ reaction rate indirectly, based on experimental results for nuclear levels in $^{30}$P, and suggested a direct measurement of the $^{29}$Si(p,$\gamma$)$^{30}$P reaction. The goal of the present work was to measure the $^{29}$Si(p,$\gamma$)$^{30}$P reaction directly, and derive improved thermonuclear rates. The astrophysical implications of our new results, together with a brief description of our experiment, have been published elsewhere \cite{downen22}, but few details were given in that work on the experimental results or the reaction rate calculation. Here, we present our detailed results on measured resonance and $\gamma$-ray energies, branching ratios, resonances strengths, and the calculation of our new $^{29}$Si(p,$\gamma$)$^{30}$P reaction rate. 

In Sec.~\ref{sec:exp}, we describe our experimental procedure. Results are presented in Sec.~\ref{sec:results}. Thermonuclear reaction rates are discussed in Sec.~\ref{sec:rates}. Comments on the nuclear structure results reported by Ref.~\cite{Lotay2020} are given in Sec.~\ref{sec:lotay}. Section~\ref{sec:summary} provides a concluding summary. 


\section{Experimental procedure}\label{sec:exp}
We measured the $^{29}$Si(p,$\gamma$)$^{30}$P reaction at the Triangle Universities Nuclear Laboratory (TUNL) using the two ion accelerators of the Laboratory for Experimental Nuclear Astrophysics (LENA). Resonances above $300$~keV bombarding energy were measured using a 1-MV (model JN) Van de Graaff accelerator, which provided beam intensities up to $40$~$\mu$A on target with $1$ $-$ $3$~keV beam energy spread. The energy calibration was established using well-known resonances in the $^{27}$Al(p,$\gamma$)$^{28}$Si reaction. Measurements below $250$~keV bombarding energy were carried out with the Electron Cyclotron Resonance Ion Source (ECRIS) accelerator. The maximum beam current on target amounted to $\approx$2.1~mA, with a beam spread of less than $1$~keV. The energy calibration was performed using the well-known $151$-keV resonance in $^{18}$O(p,$\gamma$)$^{19}$F \cite{becker1995}. For more details, see Ref.~\cite{Cooper2018}. Both accelerators have been decommissioned after the conclusion of the present experiment and will be replaced in the near future.

Gamma rays emitted from the target were analyzed with a $135$\% relative efficiency coaxial High-Purity Germanium (HPGe) detector surrounded by a $16$-segment NaI(Tl) annulus. These detectors comprise the LENA $\gamma\gamma$-coincidence spectrometer, described in more detail in Refs.~\cite{longland2006,buckner2015}. The detectors were surrounded on six sides by $1.3$-cm-thick lead panels and by five $5$-cm-thick plastic scintillator paddles used to veto spurious events induced by cosmic-ray muons. The HPGe detector was placed at a $1.1$-cm distance from the target at $0^\circ$ relative to the beam direction. Detector energy calibrations were performed using well-known room-background peaks ($^{40}$K, $^{208}$Tl) and $\gamma$-ray transitions from the $^{19}$F(p,$\alpha\gamma$)$^{16}$O reaction. 

The spectrometer is capable of reducing the room background by several orders of magnitude in the energy region below $2.6$~MeV. It is well characterized \cite{carson2010}, allowing for the determination of reliable singles and coincidence detection efficiencies in conjunction with \texttt{Geant4} simulations \cite{howard2013}. The measured pulse height spectra were modeled using a binned likelihood method with Monte Carlo simulated spectra (“templates”) \cite{dermigny20}. The fraction of the experimental spectrum belonging to each template was obtained using a Bayesian statistical approach \cite{dermigny2016}. This allowed for the extraction of the primary $\gamma$-ray branching ratios and the total number of $^{29}$Si(p,$\gamma$)$^{30}$P reactions. Corrections for coincidence summing are implicitly included in the Monte Carlo simulations used to generate the templates. This method assumes knowledge of the $\gamma$-ray branching ratios for secondary transitions in $^{30}$P, which were adopted from Refs.~\cite{Endt1990,grossmann2000}. Angular distribution corrections, when previously reported \cite{RIIHONEN1979251,Reinecke1985}, were directly implemented into the \texttt{Geant4} simulations. Two different coincidence energy gates were employed in the present work. One of the most useful gates accepted only events with $4.0$~MeV $\le$ $E_{\gamma}^{HPGe}$ $+$ $E_{\gamma}^{NaI(Tl)}$ $\le$ $6.5$~MeV for the total energy deposited in the HPGe detector and NaI(Tl) annulus. The high-energy limit excludes cosmic-ray background with energies exceeding the $^{30}$P excitation energy range of interest, while the low-energy limit excludes events caused by room background and beam-induced contaminants with relatively small $Q$-values, e.g., $^{12}$C(p,$\gamma$)$^{13}$N ($Q$ $=$ $1.94$~MeV). Only for the measurement at a bombarding energy of $227$~keV (Sec.~\ref{sec:215}) did we employ another gate, $2.9$~MeV $\le$ $E_{\gamma}^{HPGe}$ $+$ $E_{\gamma}^{NaI(Tl)}$ $\le$ $5.9$~MeV, because it improved the signal-to-background ratio in the presence of a significant beam-induced contamination from $^{19}$F(p,$\alpha\gamma$)$^{16}$O ($E_{\gamma}$ $=$ $6129$~keV).

The target was implanted using the Source of Negative Ions by Cesium Sputtering (SNICS) at the Centre de Spectrom\'etrie Nucl\'eaire et de Spectrom\'etrie de Masse (CSNSM) in Orsay, France \cite{BACRI201748}. The $^{29}$Si$^-$ beam was produced from natural silicon metalloid and implanted at $80$~keV bombarding energy into a $0.5$-mm-thick tantalum sheet. The implantation dose was $\approx 200$~mC~cm$^{-2}$. Prior to implantation, the tantalum backing was chemically etched and then outgassed in high vacuum by resistive heating to remove contaminants. The well-known resonance at $E_r^{cm}$ $=$ $403$~keV (Sec.~\ref{sec:403}) in $^{29}$Si(p,$\gamma$)$^{30}$P was used to characterize the target. The target thickness was found to be $11.4 \pm 0.3$~keV. Based on the measured maximum yield and the adopted resonance strength\footnote{The resonance strength for the $^{29}$Si $+$ $p$ reaction is defined by $\omega\gamma$ $\equiv$ $ (2J+1) \Gamma_p \Gamma_{\gamma}/(4\Gamma$), with $\Gamma_p$, $\Gamma_{\gamma}$, $\Gamma$, and $J$ denoting the proton width, $\gamma$-ray width, total width, and spin of the resonance, respectively.} (Tab.~\ref{tab:results_resonances}), the stoichiometry of the target layer was Ta/$^{29}$Si $=$ $1.2 \pm 0.2$. Yield curves measured at the end of the experiment demonstrated that both the maximum yield and thickness of the target were unchanged after an accumulated proton charge of $17$~C.

\section{Results}\label{sec:results}
\begin{figure*}[ht]
\includegraphics[width=1.5\columnwidth]{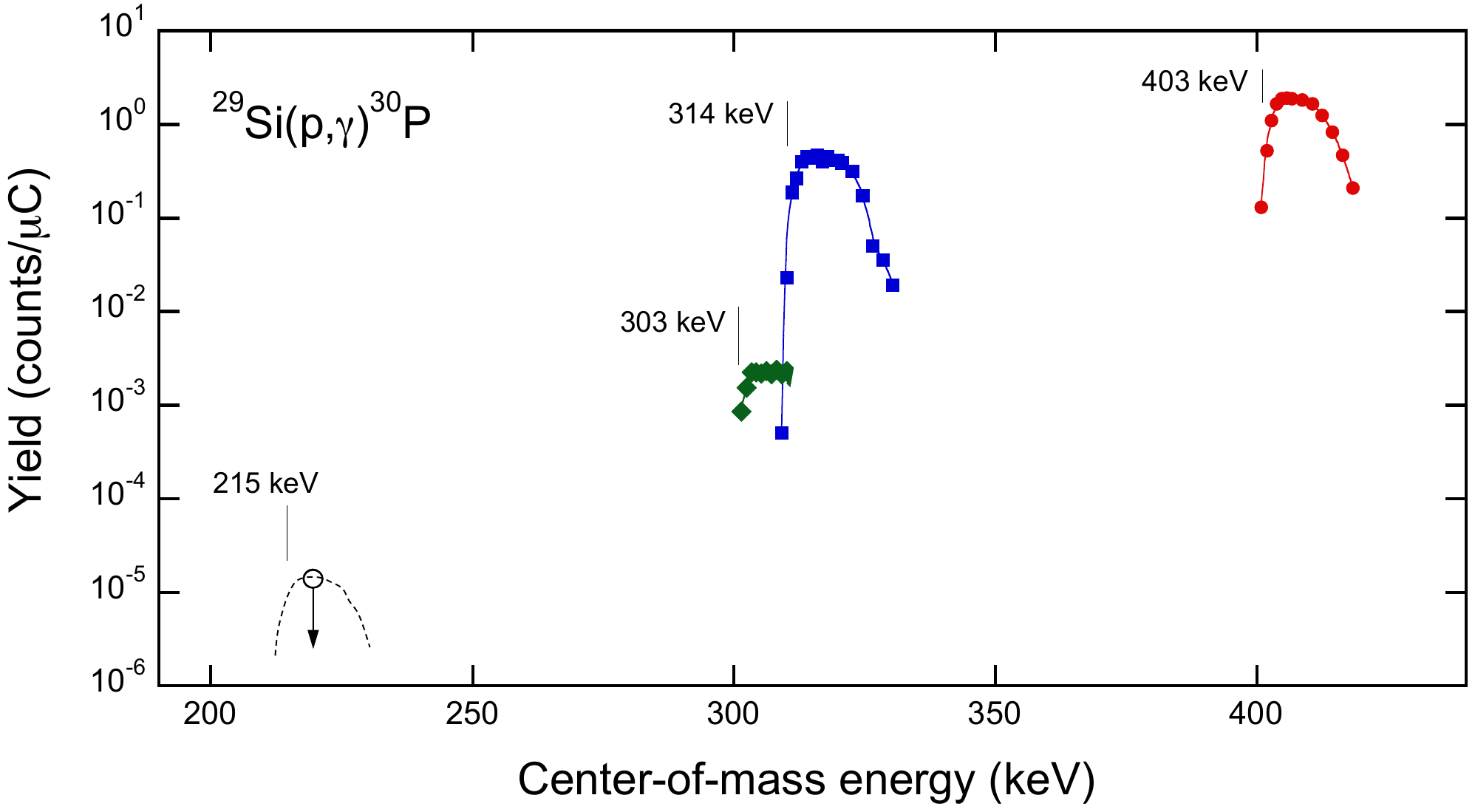}
\caption{\label{fig:yield} 
Relative yield of the $^{29}$Si(p,$\gamma$)$^{30}$P reaction versus center-of-mass energy. The depicted yields have been obtained from the singles detection mode and are not corrected for detector efficiencies. Different primary transitions are plotted for the respective resonances: 
({\color{red} $\mdsmblkcircle$}) $E_r^{cm}$ $=$ $403$~keV ($R$ $\rightarrow$ $677$~keV);
({\color{blue} $\smblksquare$}) $E_r^{cm}$ $=$ $314$~keV ($R$ $\rightarrow$ $0$~keV); 
({\color{ForestGreen} $\smblkdiamond$}) $E_r^{cm}$ $=$ $303$~keV ($R$ $\rightarrow$ $2937$~keV); ({\color{black} $\smwhtcircle$}) $E_r^{cm}$ $=$ $215$~keV ($R$ $\rightarrow$ $709$~keV). The yield for the (undetected) $215$-keV resonance represents an upper limit. The $303$-keV resonance had not been observed previously. The lines are to guide the eye only. 
}
\end{figure*}

The yield curve of $^{29}$Si(p,$\gamma$)$^{30}$P (i.e., counts in a selected primary $\gamma$-ray transition per $\mu$C of accumulated proton charge) for the four resonances investigated in the present work is displayed in Fig.~\ref{fig:yield}. Our measured $\gamma$-ray energies, $E_{\gamma}$, and branching ratios, $B_{\gamma}$, are listed in Tab.~\ref{tab:results_branch}. Branching ratios obtained from our singles and coincidence spectra (see Sec.~\ref{sec:exp}) agreed within uncertainties for each observed transition, and only the $B_{\gamma}$ values from the coincidence spectra are given in Tab.~\ref{tab:results_branch}. The weighted average of our measured $\gamma$-ray energies, after corrections for the Doppler and recoil shifts, is used to calculate the excitation energies. The lifetimes of the resonances measured here are unknown. For the calculation of the excitation energies, we assumed that the lifetimes are short compared to the slowing-down time of the recoiling $^{30}$P nuclei after capture (i.e., the observed $\gamma$-ray energies are fully Doppler shifted). Spin and parity restrictions are obtained by applying the ``Dipole or E2 rule'' \cite{Endt1990} to the observed primary $\gamma$-ray transitions. The derived excitation energies and spin and parity ($J^\pi$) ranges are listed in Tab.~\ref{tab:results_levels}, together with literature values and the properties of other states within $\approx$400~keV of the proton threshold that have not been measured in the present work. 

In the following, we will refer to the resonances using their center-of-mass energies (column 3 of Tab.~\ref{tab:results_resonances}). These have been calculated from the measured excitation energies using the Q-value based on {\it nuclear} instead of atomic masses, $Q_{nu}$ $=$ $5593.34 \pm 0.07$~keV, as discussed in Ref.~\cite{Iliadis:2019ch}. Our derived center-of-mass resonance energies and measured strengths of low-energy resonances are listed in Tab.~\ref{tab:results_resonances}. More information for each of the investigated resonances is given below. 

\begin{table*}[]
\begin{center}
\caption{Experimental $\gamma$-ray energies, $E_{\gamma}$, and branching ratios, $B_{\gamma}$, of low-energy resonances in the $^{29}$Si(p,$\gamma$)$^{30}$P reaction. All energies are in keV, branching ratios in percent. New experimental information derived in the present work is shown in boldface.\label{tab:results_branch}}
\begin{ruledtabular}
\begin{tabular}{lcccccccc}
                            & \multicolumn{3}{c}{$E_{x}$ $=$ $5996$~keV ($E_r^{cm}$ $=$ $403$~keV)} & \multicolumn{3}{c}{$E_{x}$ $=$ $5908$~keV ($E_r^{cm}$ $=$ $314$~keV)} & \multicolumn{2}{c}{$E_{x}$ $=$ $5897$~keV ($E_r^{cm}$ $=$ $303$~keV)} \\  \cline{2-4} \cline{5-7} \cline{8-9} 
Transition & $E_\gamma$\footnotemark[1]   &   $B_{\gamma}^{present}$\footnotemark[2]    & $B_{\gamma}^{previous}$\footnotemark[3] & $E_\gamma$\footnotemark[1]     &   $B_{\gamma}^{present}$\footnotemark[2]    & $B_{\gamma}^{previous}$\footnotemark[3] & $E_\gamma$\footnotemark[1]     &   $B_{\gamma}^{present}$\footnotemark[2]     \\
\hline
$R$ $\rightarrow$ $0$    ($1^+$)   &  $\bf{5994.6(14)}$    &    $\bf{3.6(4)}$      &  4.9      &  $\bf{5905.4(12)}$      &  $\bf{82.4(9)}$   &  86   &                       &      $\bf{\le 0.7}$     \\
$R$ $\rightarrow$ $677$  ($0^+$)   &  $\bf{5317.0(15)}$    &    $\bf{62.3(17)}$    &  60       &                         &                   &       &  $\bf{5217.9(20)}$    &      $\bf{1.4(2)}$      \\
$R$ $\rightarrow$ $709$  ($1^+$)   &  $\bf{5285.8(15)}$    &    $\bf{29.3(15)}$    &  30       &  $\bf{5195.2(22)}$      &  $\bf{9.2(6)}$    &  9.5  &  $\bf{5180.7(24)}$    &      $\bf{3.8(5)}$      \\
$R$ $\rightarrow$ $1454$ ($2^+$)   &                       &    $\bf{\le 0.3}$     &  0.4      &  $\bf{4452.1(19)}$      &  $\bf{5.9(4)}$    &  3.4  &                       &                         \\
$R$ $\rightarrow$ $2539$ ($3^+$)   &                       &                       &           &                         &                   &       &  $\bf{3354.4(23)}$    &      $\bf{1.8(2)}$      \\
$R$ $\rightarrow$ $2724$ ($2^+$)   &                       &                       &           &  $\bf{3183.4(24)}$      &  $\bf{0.75(6)}$   &  0.5  &                       &                         \\
$R$ $\rightarrow$ $2937$ ($2^+$)   &  $\bf{3056.9(20)}$    &    $\bf{1.8(2)}$      &  1.9      &  $\bf{2967.3(23)}$      &  $\bf{0.99(8)}$   &  0.6  &  $\bf{2960.0(12)}$    &      $\bf{69.5(15)}$    \\
$R$ $\rightarrow$ $3019$ ($1^+$)   &  $\bf{2974.7(19)}$    &    $\bf{2.1(2)}$      &  1.8      &                         &                   &       &                       &                         \\
$R$ $\rightarrow$ $3734$ ($1^+$)   &                       &                       &           &  $\bf{2174.5(18)}$      &  $\bf{0.16(3)}$   &       &                       &                         \\
$R$ $\rightarrow$ $3836$ ($2^+$)   &                       &                       &           &  $\bf{2073.0(20)}$      &  $\bf{0.29(4)}$   &       &                       &                         \\
$R$ $\rightarrow$ $4144$ ($2^-$)   &                       &                       &           &  $\bf{1764.9(19)}$      &  $\bf{0.15(2)}$   &       &                       &                         \\
$R$ $\rightarrow$ $4183$ ($2^+$)   &                       &                       &           &                         &                   &       &  $\bf{1716.1(17)}$    &       $\bf{23.5(14)}$   \\
$R$ $\rightarrow$ $4468$ ($0^+$)   &  $\bf{1527.3(16)}$    &    $\bf{0.9(1)}$      &  1.0      &                         &                   &       &                       &                         \\
$R$ $\rightarrow$ $4502$ ($1^+$)   &                       &                       &           &  $\bf{1407.2(18)}$      &  $\bf{0.16(2)}$   &       &                       &                         \\
\end{tabular}
\end{ruledtabular}
\footnotetext[1] {\footnotesize Experimental $\gamma$-ray energies; Doppler shift corrections were applied to the quoted values, but not recoil shift corrections; the common uncertainty of the $\gamma$-ray energy calibration is $\pm 0.7$~keV.}
\footnotetext[2] {\footnotesize Present values, obtained from coincidence spectra (see Sec.~\ref{sec:exp}). Upper limits correspond to 97.5\% coverage probability.}
\footnotetext[3] {\footnotesize From Ref.~\cite{Reinecke1985}, with no uncertainties reported. For the level at $E_{x}$ $=$ $5996$~keV, branching ratios have also been reported earlier by Ref.~\cite{RIIHONEN1979251}.}
\end{center}
\end{table*}

\begin{table*}[]
\begin{center}
\caption{Properties of levels near the proton threshold in $^{30}$P ($Q$ $=$ $5594.75\pm0.07$~keV \cite{wang2021}). New experimental information derived in the present work is given in boldface.\label{tab:results_levels}}
\begin{ruledtabular}
\begin{tabular}{cccccccc}
\multicolumn{2}{c}{Endt 1998 \cite{Endt1998}} & \multicolumn{2}{c}{ENSDF \cite{ensdf2021}} & \multicolumn{2}{c}{Present work} & \multicolumn{2}{c}{Adopted} \\  \cline{1-2} \cline{3-4} \cline{5-6} \cline{7-8}
$E_x$ (keV) & $J^\pi$  &  $E_x$ (keV) & $J^\pi$  & $E_x$ (keV)\footnotemark[6] & $J^\pi$\footnotemark[7]  & $E_x$ (keV) & $J^\pi$    \\
\hline
$5595 \pm 3$                        &   $4^+$                       &   $5597 \pm 5$                        &   $4^+$                       &                           &                   & $5597 \pm 5$                          &   $4^+$ \\
$5701.7 \pm 0.4$\footnotemark[1]    &   $1^+$                       &   $5701.3 \pm 0.2$                    &   $1^+$                       &                           &                   & $5701.3 \pm 0.2$                      &   $1^+$ \\ 
$5714 \pm 3$                        &   $(5,7)^+$                   &   $5715 \pm 4$                        &   $(5,7)^+$                   &                           &                   & $5715 \pm 4$                          &   $(5,7)^+$ \\
                                    &                               &   $5788 \pm 5$\footnotemark[2]        &   $(3-5)^+$                   &                           &                   & $5788 \pm 5$                          &   $(3-5)^+$ \\
$5808 \pm 3$\footnotemark[10]       &   $(3,5)^+$                   &   $5808 \pm 5$                        &   $(3,5)^+$                   &                           &                   & $5808 \pm 3$\footnotemark[10]         &   $3^+$\footnotemark[9] \\
$5890 \pm 12$\footnotemark[4]       &   $(1-3)^+$\footnotemark[4]   &   $5896 \pm 5$\footnotemark[5]        &   $(2^-)$\footnotemark[5]     & $\bf{5896.7 \pm 1.0}$     & $\bf{(1,2)^+}$    & $\bf{5896.7 \pm 1.0}$                 &   $\bf{(1,2)^+}$ \\            
$5907.7 \pm 0.8$\footnotemark[3]    &   $2^-$                       &   $5907.8 \pm 0.8$\footnotemark[3]    &   $(2,1)^-$                   & $\bf{5907.2 \pm 0.9}$     & $\bf{(1,2,3^+)}$  & $\bf{5907.5 \pm 0.6}$\footnotemark[8] &   $(2,1)^-$ \\
$5934.0 \pm 0.5$                    &                               &   $5934.0 \pm 0.1$                    &   $(3^+)$                     &                           &                   & $5934.0 \pm 0.1$                      &   $(3^+)$ \\ 
$5993 \pm 4$                        &   $(0-2)^-$                   &   $5993 \pm 5$                        &   $(0-2)^-$                   &                           &                   & $5993 \pm 5$                          &   $(0-2)^-$ \\
$5997.1 \pm 0.8$\footnotemark[3]    &   $1^+$                       &   $5997.2 \pm 0.8$                    &   $(1^+)$                     &  $\bf{5995.0 \pm 0.9}$    &  $\bf{(1,2^+)}$   & $\bf{5996.2 \pm 0.6}$\footnotemark[8] &   $(1^+)$ \\
\end{tabular}
\end{ruledtabular}
\footnotetext[1] {\footnotesize Mean lifetime $\tau_m$ $=$ $16 \pm 4$~fs \cite{Tikkanen1987}. See Sec~\ref{sec:5701}.}
\footnotetext[2] {\footnotesize This level, which is not listed in Endt \cite{Endt1998}, is clearly populated as part of a doublet ($E_x$ $=$ $5788$~keV and $5807$~keV) in the $^{30}$Si($^3$He,t)$^{30}$P study of Ref.~\cite{Ramstein1981}, where an $L$ $=$ $4$ transfer is suggested, resulting in $J^\pi$ $=$ $(3-5)^+$.}
\footnotetext[3] {\footnotesize The excitation energy is calculated from the directly-measured resonance energy \cite{Endt1998}.}
\footnotetext[4] {\footnotesize Populated in the $^{31}$P($^3$He,$\alpha$)$^{30}$P experiment of Ref.~\cite{vanGasteren1974}, who report an angular momentum transfer of $\ell_n$ $=$ $2$, resulting in a tentative spin-parity assignment of $(1-3)^+$. The level energy is reported in ENSDF \cite{ensdf2021}, but is omitted from their main table.}
\footnotetext[5] {\footnotesize Populated in the $^{30}$Si($^3$He,t)$^{30}$P experiment of Ref.~\cite{Ramstein1981}, who report an angular momentum transfer of $L$ $=$ $3$ and an unambiguous $2^-$ assignment. See discussion in Sec.~\ref{sec:303}.}
\footnotetext[6] {\footnotesize Weighted average from the measured $\gamma$-ray energies (Tab.~\ref{tab:results_branch}), assuming they are fully Doppler shifted (Sec.~\ref{sec:results}).}
\footnotetext[7] {\footnotesize From the application of the ``Dipole or E2 rule'' \cite{Endt1990} to the observed transitions (Tab.~\ref{tab:results_branch}).}
\footnotetext[8] {\footnotesize Weighted average from Ref.~\cite{Endt1998} and present work.}
\footnotetext[9] {\footnotesize Combined assignment using $J^\pi$ $=$ $(3-5)^+$ \cite{Ramstein1981}, $J^\pi$ $=$ $(1-3)^+$ \cite{Dykoski1976}, and $J^\pi$ $=$ $U$ (unnatural parity) \cite{Boerma1975}. See Sec.~\ref{sec:215}.}
\footnotetext[10] {\footnotesize Weighted average of values measured by Refs.~\cite{Hafner1974,Boerma1975,Dykoski1976,Ramstein1981}.}
\end{center}
\end{table*}

\begin{table*}[]
\begin{center}
\caption{Properties of low-energy resonances in the $^{29}$Si(p,$\gamma$)$^{30}$P reaction. New experimental information derived in the present work is shown in boldface.\label{tab:results_resonances}}
\begin{ruledtabular}
\begin{tabular}{cccccc}
$E_x$ (keV)\footnotemark[1] & $J^\pi$\footnotemark[1]  &  $E_r^{cm}$ (keV)\footnotemark[2] & $\omega\gamma^{present}$ (eV) & $\omega\gamma^{previous}$ (eV) & $\Gamma_p^{present}$ (eV)\footnotemark[6] \\
\hline
$5597 \pm 5$            &   $4^+$           &  $3.7 \pm 5.0$            &                                                   &                                   & $\le$ $8.5 \times 10^{-97}$\footnotemark[7] \\
$5701.3 \pm 0.2$        &   $1^+$           &  $108.0 \pm 0.2$          &                                                   &                                   & $\le$ $1.2 \times 10^{-10}$ \\ 
$5715 \pm 4$            &   $(5,7)^+$       &  $121.7 \pm 4.0$          &                                                   &                                   & $\le$ $3.0 \times 10^{-16}$ \\ 
$5788 \pm 5$            &   $(3-5)^+$       &  $194.7 \pm 5.0$          &                                                   &                                   & $\le$ $1.5 \times 10^{-8}$ \\ 
$5808 \pm 3$            &   $3^+$           &  $214.7 \pm 3.0$          &   $\bf{\le 3.3 \times 10^{-7}}$\footnotemark[5]   &                                   & $\approx 1.0 \times 10^{-7}$ \\ 
$\bf{5896.7 \pm 1.0}$   &   $\bf{(1,2)^+}$  &  $\bf{303.4 \pm 1.0}$     &   $\bf{(8.8 \pm 1.5)\times 10^{-5}}$              &                                   & \\            
$\bf{5907.5 \pm 0.6}$   &   $(2,1)^-$       &  $\bf{314.2 \pm 0.6}$     &   $\bf{0.0207 \pm 0.0027}$                        & $0.015 \pm 0.005$\footnotemark[3] & \\ 
$5934.0 \pm 0.1$        &   $(3^+)$         &  $340.7 \pm 0.1$          &                                                   &                                   & \\ 
$5993 \pm 5$            &   $(0-2)^-$       &  $399.7 \pm 5.0$          &                                                   &                                   & \\ 
$\bf{5996.2 \pm 0.6}$   &   $(1^+)$         &  $\bf{402.9 \pm 0.6}$     &                                                   & $0.220 \pm 0.025$\footnotemark[4] & \\ 
\end{tabular}
\end{ruledtabular}
\footnotetext[1] {\footnotesize From columns 7 and 8 of Tab.~\ref{tab:results_levels}.}
\footnotetext[2] {\footnotesize Calculated from column 1 using the Q-value based on {\it nuclear} masses: $Q_{nu}$ $=$ $5593.34 \pm 0.07$~keV \cite{Iliadis:2019ch}.}
\footnotetext[3] {\footnotesize From Ref.~\cite{Reinecke1985}; earlier values: $\omega\gamma$ $=$ $0.019 \pm 0.004$~eV \cite{Harris1969}, $0.0175 \pm 0.0050$~eV \cite{PhysRev.110.96}.}
\footnotetext[4] {\footnotesize From Ref.~\cite{SARGOOD198261} (see Sec.~\ref{sec:results}).}
\footnotetext[5] {\footnotesize Corresponding to 97.5\% coverage probability.}
\footnotetext[6] {\footnotesize Indirect estimate using Eq.~(\ref{eq:gamma}) and proton spectroscopic factors derived from Ref.~\cite{Dykoski1976}. Since $\Gamma_p$ $\ll$ $\Gamma_{\gamma}$ for low-energy resonances, it follows that $\omega\gamma$ $\approx$ $(2J+1) \Gamma_p/4$. See Sec.~\ref{sec:unobserved}.}
\footnotetext[7] {\footnotesize Disregarded for the calculation of the total reaction rate; see Sec.~\ref{sec:5597}.}
\end{center}
\end{table*}
%

%

\subsection{$E_r^{cm}$ $=$ $403$~keV, $E_x$ $=$ $5996$~keV [$J^\pi$ $=$ $(1^+)$]}\label{sec:403}
The yield curve of the $E_r^{cm}$ $=$ $403$~keV resonance for the strongest primary transition, $R$ $\rightarrow$ $677$~keV, is shown in red in Fig.~\ref{fig:yield}. Measured $\gamma$-ray energies and branching ratios are given in columns 2 and 3, respectively, of Tab.~\ref{tab:results_branch}. The branching ratios have been corrected for angular distribution effects using the coefficients measured by Ref.~\cite{RIIHONEN1979251}. Branching ratios reported by Reinecke {\it et al.} \cite{Reinecke1985} are listed for comparison. Although Ref.~\cite{Reinecke1985} does not report any uncertainties, the previous and present branching ratios are in overall agreement. The only exception is the weak branch to the $E_x$ $=$ $1454$~keV ($2^+$) level. Reinecke {\it et al.} \cite{Reinecke1985} reported a value of $B_{\gamma}$ $=$ $0.4$\%, but it is neither observed in our singles nor coincidence spectra. Our deduced upper limit for this branching is $\le0.3$\% ($97.5$\% coverage probability). 

Values for the excitation energy were derived from the measured $\gamma$-ray energies. The weighted average resulting from the six observed transitions is $E_x^{present}$ $=$ $5995.0 \pm 0.9$~keV. Averaging our value with the previous result \cite{Endt1998} yields a recommended excitation energy of $E_x^{adopted}$ $=$ $5996.2 \pm 0.6$~keV (Tab.~\ref{tab:results_levels}). From the adopted excitation energy, we find a center-of-mass resonance energy of $E_r$ $=$ $402.9 \pm 0.6$~keV (Tab.~\ref{tab:results_resonances}). Our estimated range for the spin-parity of the compound level, $J^\pi$ $=$ $(1,2^+)$, is wider than, but in agreement with, the previous estimate, $(1^+)$. 

Resonance strength values have been reported by Refs.~\cite{BROUDE19561139,PhysRev.110.96,ENGELBERTINK196612,RIIHONEN1979251}, with differences amounting to a factor of $\approx$4.5. In the present work, we will adopt the strength value listed in Tab.~1 of Sargood~\cite{SARGOOD198261}, $\omega\gamma$ $=$ $0.220 \pm 0.025$~eV (Tab.~\ref{tab:results_resonances}), which is consistent with the set of standard (p,$\gamma$) strengths used in nuclear astrophysics (see, e.g., Refs.~\cite{PAINE1979389,Iliadis2001,Iliadis_2015}). We adopted this strength as a standard for determining the strengths, or upper limits, of the other resonances measured in the present work, and also to estimate the target stoichiometry (Sec.~\ref{sec:exp}). 

\subsection{$E_r^{cm}$ $=$ $314$~keV, $E_x$ $=$ $5908$~keV [$J^\pi$ $=$ $(2,1)^-$]}\label{sec:314}
The yield of the $E_r^{cm}$ $=$ $314$~keV resonance for the strongest primary transition, $R$ $\rightarrow$ $0$~keV, is displayed in blue in Fig.~\ref{fig:yield}. Values for the excitation energy were derived from our measured $\gamma$-ray energies (Tab.~\ref{tab:results_branch}). The weighted average resulting from the nine observed transitions is $E_x^{present}$ $=$ $5907.2 \pm 0.9$~keV. Averaging our value with the previous result \cite{Endt1998} yields a recommended excitation energy of $E_x^{adopted}$ $=$ $5907.5 \pm 0.6$~keV (Tab.~\ref{tab:results_levels}). From the adopted excitation energy, we find a center-of-mass resonance energy of $E_r^{cm}$ $=$ $314.2 \pm 0.6$~keV (Tab.~\ref{tab:results_resonances}).

The branching ratios (Tab.~\ref{tab:results_branch}) have been corrected for angular distribution effects using the coefficients measured by Reinecke {\it et al.} \cite{Reinecke1985}. Branching ratios reported by Ref.~\cite{Reinecke1985} are listed for comparison. We observed four new primary branches, to the levels at $E_x$ $=$ $3734$~keV, $3836$~keV, $4144$~keV, and $4502$~keV, in both the singles and coincidence spectra. Our estimated range for the spin-parity of the compound level is $J^\pi$ $=$ $(1, 2, 3^+)$, in agreement with the more restricted range, $(2,1)^-$, reported in ENSDF \cite{ensdf2021} (Tab.~\ref{tab:results_levels}).

Our result for the resonance strength is $\omega\gamma^{present}$ $=$ $0.0207 \pm 0.0027$~eV, representing a $13$\% uncertainty. Most of the uncertainty derives from that of the standard resonance at $E_r^{cm}$ $=$ $403$~keV ($11$\%). Other sources of relative systematic uncertainty are the experimental detector efficiency ($5.0$\%; including the \texttt{Geant4} simulations), beam charge integration ($3.0$\%), and stopping powers ($5.0$\%). The statistical uncertainty is less than $1.0$\%. Our measured resonance-strength value agrees within $1\sigma$ with previous results \cite{PhysRev.110.96,Harris1969,Reinecke1985} (see Tab.~\ref{tab:results_resonances}), but our uncertainty is smaller by a factor of $\approx 2$. 

Note that the previous evaluation \cite{LONGLAND2010,ILIADIS2010b,ILIADIS2010c,iliadis2010d} of the $^{29}$Si(p,$\gamma$)$^{30}$P rate assumed a value of $\omega\gamma^{2010}$ $=$ $0.0127 \pm 0.0042$~eV, which differs significantly from the present result. The $2010$ value was obtained by adopting the strength of Reinecke {\it et al.} \cite{Reinecke1985} (see Tab.~\ref{tab:results_resonances}; also listed in Ref.~\cite{Endt1990}) and renormalizing it to the standard strength of Sargood~\cite{SARGOOD198261}.

\subsection{$E_r^{cm}$ $=$ $303$~keV, $E_x$ $=$ $5897$~keV [$J^\pi$ $=$ $(1,2)^+$]}\label{sec:303}
\begin{figure*}[ht]
\includegraphics[width=1.7\columnwidth]{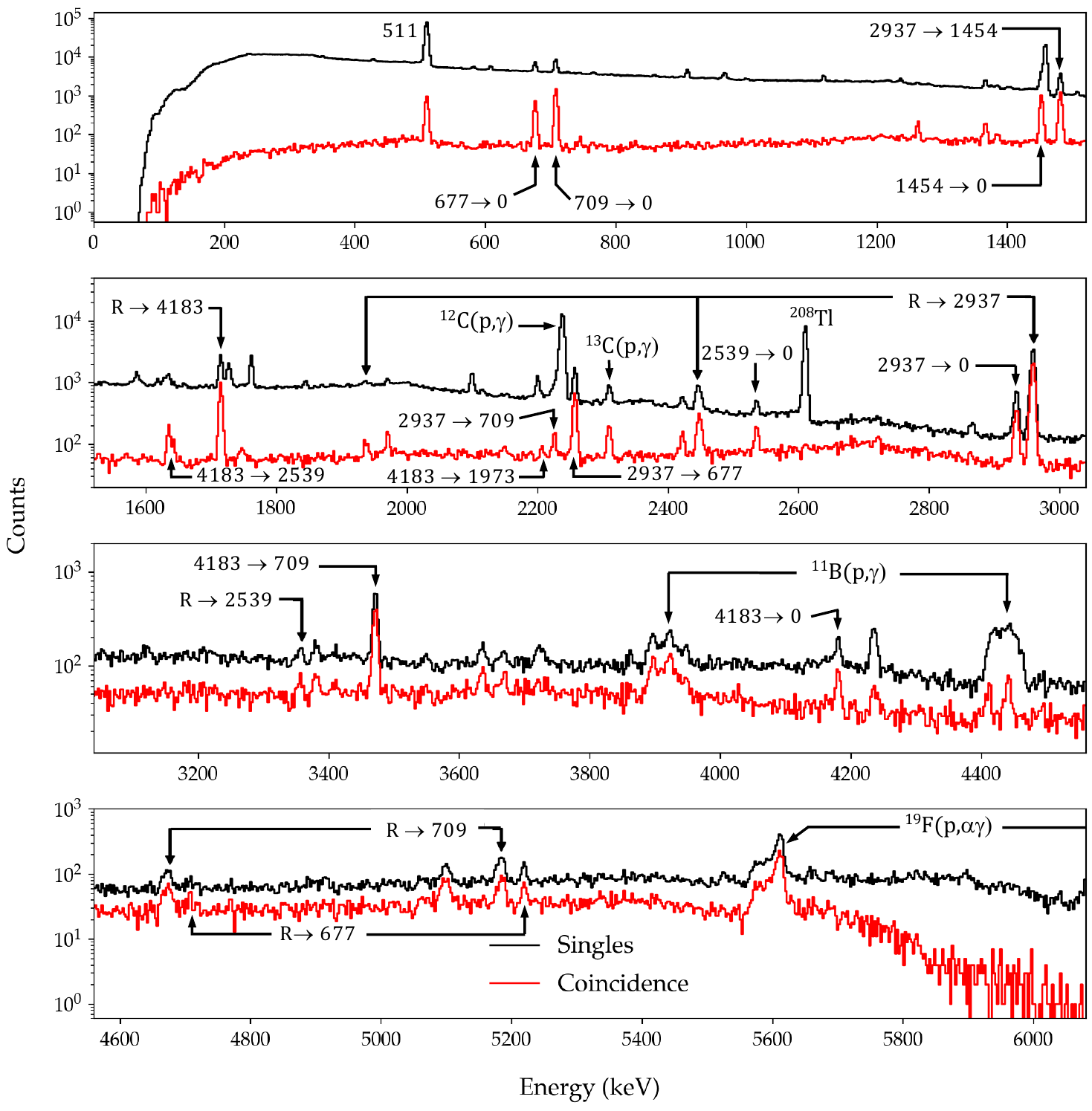}
\caption{\label{fig:spec} 
Singles (black) and coincidence (red) spectra for the previously unobserved $E_r^{cm}$ $=$ $303$~keV resonance ($E_x$ $=$ $5897$~keV) in $^{29}$Si(p,$\gamma$)$^{30}$P, measured at a center-of mass energy of $306$~keV. Primary and secondary transitions are labeled by ``$R$ $\rightarrow$ $x$'' and ``$y$ $\rightarrow$ $x$'', respectively, where $x$ and $y$ are excitation energies in units of keV. The spectra are not contaminated by the nearby resonance at $E_r^{cm}$ $=$ $314$~keV (Sec.~\ref{sec:314}), because its strongest branching ($R$ $\rightarrow$ $0$~keV; see Tab.~\ref{tab:results_branch}) is neither observed in the singles spectrum (at the peak energy of $E_{\gamma}$ $=$ $5910$~keV), nor in the coincidence spectrum (at the single-escape peak energy of $E_{\gamma}^{\prime}$ $=$ $5400$~keV)).
}
\end{figure*}

The yield of a previously unobserved resonance at $E_r^{cm}$ $=$ $303$~keV for the strongest primary transition, $R$ $\rightarrow$ $2937$~keV, is shown in green in Fig.~\ref{fig:yield}. Singles (black) and coincidence (red) spectra of this resonance, measured at a center-of mass energy of $306$~keV, are presented in Fig.~\ref{fig:spec}. Altogether we observed five different primary transitions, both in the singles and the coincidence spectra. Inspection of Tab.~\ref{tab:results_branch} demonstrates that the $E_r^{cm}$ $=$ $303$~keV resonance is distinct from the $E_r^{cm}$ $=$ $314$~keV resonance (Sec.~\ref{sec:314}) because the respective branching ratios differ significantly. Gamma-ray energies and peak intensities were derived from spectra recorded below the $E_r^{cm}$ $=$ $314$~keV resonance, where the yield of the latter is negligible. For example, it can be seen in Tab.~\ref{tab:results_branch} that only an upper limit ($\le 0.7\%$) is measured for the ground-state branch of the $E_r^{cm}$ $=$ $303$~keV resonance, although this is the dominant primary branch of the $E_r^{cm}$ $=$ $314$~keV resonance. Furthermore, from the observed primary transitions, we estimated a spin-parity range of $J^\pi$ $=$ $(1, 2)^+$ for this new resonance.

Values for the excitation energy were derived from the measured $\gamma$-ray energies (Tab.~\ref{tab:results_branch}) of the five observed primary transitions. The result is $E_x^{present}$ $=$ $5896.7 \pm 1.0$~keV (Tab.~\ref{tab:results_levels}), corresponding to a center-of-mass resonance energy of $E_r^{cm}$ $=$ $303.4 \pm 1.0$~keV (Tab.~\ref{tab:results_resonances}). 

The information available in the literature on energies and $J^\pi$ values near this energy is ambiguous. The excitation energy of $5890 \pm 12$~keV listed in Endt \cite{Endt1990} was adopted from the $^{31}$P($^3$He,$\alpha$)$^{30}$P experiment of van Gasteren {\it et al.} \cite{vanGasteren1974}. They reported an angular momentum transfer\footnote{We use the symbol $\ell_x$ for the orbital angular momentum in the transfer (pickup or stripping) of a single nucleon ($x$ $=$ $n$: neutron; $x$ $=$ $p$: proton) and reserve the symbol $L$ for the transferred orbital angular momentum of a composite particle ($d$, $t$, $\alpha$, etc.)  or in a charge-exchange reaction \cite{Endt1990}.} of $\ell_n$ $=$ $2$, resulting in a spin-parity range of $(1-3)^+$. This level energy is listed in ENSDF \cite{ensdf2021}, but is omitted from their main table. The excitation energy evaluated in ENSDF ($5896$~keV) was adopted from the $^{30}$Si($^3$He,t)$^{30}$P experiment of Ramstein {\it et al.} \cite{Ramstein1981}, who reported an angular momentum transfer of $L$ $=$ $3$ and an unambiguous $2^-$ assignment. However, Ref.~\cite{Ramstein1981} only present in their Fig.~10 the combined angular distribution fit for the doublet ($5896$~keV and $5928$~keV), which shows little structure. Their reported unambiguous assignment resulted from their assumption that {\it ``...[the 5896~keV] level probably corresponds to the $E_x$ $=$ $5910 \pm 5$~keV; $J^\pi$ $=$ $(1,2)^-$ state. The combined information would then lead to $J^\pi$ $=$ $2^-$ for this state.''} If the assumption of Ref.~\cite{Ramstein1981} is correct, the level should not have been listed in ENSDF as separate from the $5908$-keV state (see Tab.~\ref{tab:results_levels}). On the other hand, if this assumption is incorrect, and the $5896$-keV level is distinct from the $E_x$ $=$ $5908$~keV state, the assigning of a $L$ $=$ $3$ transfer to the $5896$-keV component of the doublet is likely erroneous because it implies a parity opposite to that determined by Ref.~\cite{vanGasteren1974} and in the present work. For these reasons, we will disregard the value of $5896 \pm 5$~keV \cite{Ramstein1981} and adopt the present result for the recommended level energy (see Tab.~\ref{tab:results_levels}).

For the resonance strength, we find $\omega\gamma$ $=$ $(8.8 \pm 1.5) \times 10^{-5}$~eV, representing a $17$\% uncertainty (Tab.~\ref{tab:results_resonances}). For information on systematic uncertainties, see Sec.~\ref{sec:314}. In the 2010 thermonuclear rate evaluation \cite{LONGLAND2010,ILIADIS2010b,ILIADIS2010c,iliadis2010d}, this resonance had an energy of $E_{r,2010}^{cm}$ $=$ $296 \pm 12$~keV and a strength of $\omega\gamma^{2010}$ $\approx$ $4 \times 10^{-5}$~eV. The latter order-of-magnitude estimate was found by using the measured proton spectroscopic factor of the analog level in $^{30}$Si. 

%

\subsection{$E_r^{cm}$ $=$ $215$~keV, $E_x$ $=$ $5808$~keV [$J^\pi$ $=$ $3^+$]}\label{sec:215}
The information on the $^{30}$P level structure near $E_x$ $\approx$ $5800$~keV excitation that is presented in the literature is ambiguous (Tab.~\ref{tab:results_levels}). It appears that the level at $5808$~keV, observed in the $^{30}$Si($^3$He,t)$^{30}$P experiment of Ref.~\cite{Ramstein1981}, has been disregarded in the compilation of Endt \cite{Endt1998}, although it is clearly seen as one component of a doublet ($E_x$ $=$ $5788$ and $5808$~keV). Both of these levels are listed in ENSDF \cite{ensdf2021}. The ($^3$He,$t$) angular distribution measured by Ramstein {\it et al.} \cite{Ramstein1981} suggests $L$ $=$ $4$ for both states, implying spin-parity ranges of $J^\pi$ $=$ $(3-5)^+$. For the $5808$-keV component of the doublet, two more spin-parity restrictions are available: $J^\pi$ $=$ $(1-3)^+$ from the observation of a $\ell_p$ $=$ $2$ transfer in the $^{29}$Si($^3$He,d)$^{30}$P experiment of Ref.~\cite{Dykoski1976}, and an unnatural parity assignment from the $^{32}$S($\vec{d}$,$\alpha$)$^{30}$P study of Ref.~\cite{Boerma1975}. The resulting assignment is $J^\pi$ $=$ $3^+$. However, the $5808$-keV level is listed with $J^\pi$ $=$ $(3,5)^+$ in both Endt \cite{Endt1998} and ENSDF \cite{ensdf2021}. It appears that Ref.~\cite{Endt1998} has disregarded the assignment from the ($^3$He,d) study. Furthermore, Ref.~\cite{ensdf2021} states that their spin-parity assignment results from the ($^3$He,d) and ($^3$He,t) studies, which, however, would have resulted in $J^\pi$ $=$ $3^+$. We will disregard here the ``$J^\pi$ $=$ $5^+$'' assignment of Ref.~\cite{grossmann2000} since it is entirely based on a comparison with shell-model calculations. For the level energy and uncertainty, we adopted Endt's result ($5808 \pm 3$~keV; Tab.~\ref{tab:results_levels}), which was obtained from the weighted average of the values measured by Refs.~\cite{Hafner1974,Boerma1975,Dykoski1976,Ramstein1981}. 

We searched for this resonance by measuring singles and coincidence spectra with an accumulated beam charge of $\approx$10~C. For the laboratory bombarding energy, we chose $227$~keV, which is near the center of the expected yield curve appropriate for a target thickness of $\approx$14~keV at this bombarding energy. 

We did not observe any primary or secondary transitions in $^{30}$P in either the singles or coincidence spectra. Therefore, only an upper limit on the resonance strength can be obtained. To estimate this upper limit, we proceeded as follows. Although two decays of this level have been reported in Ref.~\cite{Lotay2020}, the primary branching ratios are unknown. To derive a conservative upper limit and not rely on indefensible assumptions, we assumed that the primary decays will proceed either to the ground state or to any of seven excited levels below $3$-MeV excitation energy. We excluded the primary decay to the $677$-keV ($0^+$) level from consideration because it would imply an unlikely $M3$ transition. Templates were simulated for these decays using \texttt{Geant4} (Sec.~\ref{sec:exp}), and the measured spectra were fitted with these templates using the same Bayesian approach employed in the analysis of the other resonances discussed above. The resonance strength estimate was obtained from the posterior of the total number of $^{29}$Si $+$ $p$ reactions. By applying corrections for experimental artifacts (see Sec.~\ref{sec:314}), we obtained an upper limit of $\omega\gamma$ $\le$ $3.3 \times 10^{-7}$~eV (97.5\% coverage probability). Possible angular distribution effects are well contained in the upper limit value.

The derived value was significantly impacted by beam-induced background in our implanted target from the $E_r^{lab}$ $=$ $224$~keV resonance in $^{19}$F(p,$\alpha\gamma$)$^{16}$O. Although we could reduce this background by adjusting the energy thresholds of the coincidence gates (Sec.~\ref{sec:exp}), it is not possible to remove it because the energy of the contaminant $\gamma$ rays (e.g., $E_{\gamma}$ $=$ $6129$~keV) exceeds that for the $^{30}$P transitions of interest. The $E_r^{cm}$ $=$ $215$~keV resonance will be discussed in more detail in Sec.~\ref{sec:5808}.

\section{Thermonuclear reaction rates}\label{sec:rates}
The reaction rate for $^{29}$Si($p$,$\gamma$)$^{30}$P depends on both resonant and non-resonant contributions. The total thermonuclear rate (in units of cm$^3$ mol$^{-1}$ s$^{-1}$) for a reaction involving two nuclei (0 and 1) in the entrance channel at a given temperature is given by \cite{Iliadis_2015}
\begin{align}
    N_A \langle \sigma v \rangle = \frac{ 3.7318 \times 10^{10}}{T_9^{3/2}} \sqrt{\frac{M_0 + M_1}{M_0 M_1}} \nonumber \\
    \times \int_0^{\infty} E \sigma(E) e^{-11.605 E /T_9} dE
\end{align}
where the center-of-mass energy, $E$, is in units of MeV, the temperature, $T_9$, is in GK ($T_9 \equiv T/10^9$ K), the nuclear masses, $M_i$, are in u, the cross section, $\sigma$, is in barn ($1$ b $\equiv 10^{-24}$ cm$^2$), and $N_A$ denotes Avogadro's constant. We estimated experimental thermonuclear rates for the $^{29}$Si(p,$\gamma$)$^{30}$P reaction using the Monte Carlo procedure presented in Ref.~\cite{LONGLAND2010,ILIADIS2010b,ILIADIS2010c,iliadis2010d}, which fully implements the uncertainties of all measured input quantities (e.g., resonance energies, strengths, partial widths, non-resonant S-factors). To perform the Monte Carlo sampling, we used the program \texttt{RatesMC} \cite{LONGLAND2010}, which computes the probability density function of the total reaction rate on a fixed temperature grid. At each temperature, 20,000 samples were drawn. Low, recommended, and high rates were obtained from the 16, 50, and 84 percentiles, respectively, of the probability density function. Information on partial reaction rate contributions will be discussed next, before presenting the total rates.

\subsection{Direct capture}\label{sec:dc}
Direct proton capture has not been observed in the $^{29}$Si($p$,$\gamma$)$^{30}$P reaction, but it is nevertheless expected to contribute to the total reaction rate, especially at low temperatures \cite{Iliadis2001}. An estimate of its contribution can be obtained by calculating the direct capture cross section for each transition to a $^{30}$P bound state using a potential model and weighing the result by the measured proton spectroscopic factor \cite{Dykoski1976}. The sum over all bound states gives the total direct capture cross section. For more information about this procedure, see Ref.~\cite{Iliadis:2004ej}. Converting the direct capture cross section into an astrophysical S-factor and presenting the result numerically as a Taylor series gives
\begin{align}
S_{DC}(E) = 0.1072 - 1.262 \times 10^{-2} E +\frac{1}{2} 1.114 \times 10^{-3} E^2 
\end{align}
where $E$ and $S_{DC}$ are in units of MeV and MeVb, respectively. For the uncertainty of the $S$-factor, we assumed 40\%.

\subsection{Resonances at $E_r^{cm}$ $=$ $303$, $314$, and $403$~keV}\label{sec:ratereslow}
For the lowest-lying directly measured resonances, we used energy and strength values that incorporate new information obtained in the present work (Tabs.~\ref{tab:results_levels} and \ref{tab:results_resonances}). Recall that we did not independently measure the strength of the $E_r^{cm}$ $=$ $403$~keV resonance, but adopted the value of Sargood \cite{SARGOOD198261} as a standard for our measured strengths (Sec.~\ref{sec:403}).

\subsection{Previously measured resonances at $E_r^{cm}$ $=$ $675$ $-$ $3076$~keV}\label{sec:ratereshigh}
Energies and strengths of previously known resonances in the center-of-mass energy range of $E_r^{cm}$ $=$ $675$ $-$ $3076$~keV are compiled in Tab.~30f of Ref.~\cite{Endt1998}. We adopted these values with two modifications. First, the strengths listed in Ref.~\cite{Endt1998} are normalized to the $E_r^{cm}$ $=$ $403$~keV resonance measured by Riihonen {\it et al.} \cite{RIIHONEN1979251}. We renormalized all strengths using the value recommended for this resonance by Sargood \cite{SARGOOD198261} (see Sec.~\ref{sec:403}). Second, we calculated the center-of-mass resonance energies from the listed excitation energies using the $Q$ value derived from {\it nuclear} masses, as discussed in Ref.~\cite{Iliadis:2019ch} (Sec.~\ref{sec:results}). For a number of resonances at the higher energies, no uncertainties are listed for the resonance strengths in Ref.~\cite{Endt1998}. In those cases, we assumed an uncertainty of 25\%.  

\subsection{Unobserved resonances at $E_r^{cm}$ $\le$ $300$~keV}\label{sec:unobserved}
Five levels are known just above the proton threshold in $^{30}$P (Tabs.~\ref{tab:results_levels} and \ref{tab:results_resonances}) that have not been observed as resonances, but that nevertheless could contribute to the total reaction rate. These are discussed below.

\subsubsection{$E_r^{cm}$ $=$ $215$~keV, $E_x$ $=$ $5808$~keV [$J^\pi$ $=$ $3^+$]}\label{sec:5808}
This level, corresponding to a resonance energy of $E_r^{cm}$ $=$ $215$~keV, has already been discussed in Sec.~\ref{sec:215}. It has been observed in the reactions $^{30}$Si($^3$He,t)$^{30}$P \cite{Ramstein1981}, $^{29}$Si($^3$He,d)$^{30}$P \cite{Dykoski1976}, $^{28}$Si($^3$He,p)$^{30}$P \cite{Hafner1974}, and $^{32}$S($\vec{d}$,$\alpha$)$^{30}$P \cite{Boerma1975}, and we reported in Sec.~\ref{sec:215} an experimental upper limit on its resonance strength (Tab.~\ref{tab:results_resonances}).

From the information provided in the proton-transfer study of Dykoski {\it et al.} \cite{Dykoski1976}, we can also estimate the proton partial width indirectly. The level is clearly populated in the deuteron spectrum of their Fig.~2, and the measured angular distribution is depicted in their Fig.~7. They reported a proton spectroscopic factor of $C^2S$ $=$ $0.009$ (for a $\ell_p$ $=$ $2$ transfer and $J^\pi$ $=$ $3^+$). We calculated the proton partial width from \cite{LONGLAND2010} 
\begin{align}\label{eq:gamma}
\Gamma_p = 2 \frac{\hbar^2}{m R^2} P_c C^2S~\theta_{sp}^2
\end{align}
with $m$ being the reduced mass, $R$ the channel radius, $P_c$ the penetration factor, and $\theta_{sp}^2$ the dimensionless single-particle reduced with. The value for the latter quantity was adopted from Ref.~\cite{ILIADIS1997}. The result for the indirect estimate of the proton partial width is $\Gamma_p^{est}$ $\approx$ $2.0 \times 10^{-7}$~eV. Note that our measured resonance strength upper limit (Tab.~\ref{tab:results_resonances}) corresponds to a proton partial width upper limit of $\Gamma_p^{exp}$ $\le$ $1.9 \times 10^{-7}$~eV. We will combine both results and adopt for the proton partial width a median value of $\Gamma_p^{215}$ $=$ $1.0 \times 10^{-7}$~eV with an uncertainty of a factor of $2$ (Tab.~\ref{tab:results_resonances}).

The assumptions discussed above differ from those adopted in the 2010 thermonuclear rate evaluation \cite{LONGLAND2010,ILIADIS2010b,ILIADIS2010c,iliadis2010d}. At the time, a unique spin-parity was not listed in the evaluation of Ref.~\cite{Endt1998}. Since a possible $5^+$ assignment would have reduced the estimate of the proton partial width, only an upper limit for the proton partial width was adopted previously.

\subsubsection{$E_r^{cm}$ $=$ $195$~keV, $E_x$ $=$ $5788$~keV [$J^\pi$ $=$ $(3-5)^+$]}\label{sec:5788}
The level at $E_x$ $=$ $5788$~keV, corresponding to a resonance energy of $E_r^{cm}$ $=$ $195$~keV, has not been taken into account in previous $^{29}$Si $+$ $p$ rate evaluations because it was not mentioned by Endt \cite{Endt1998}. However, it is listed in ENSDF \cite{ensdf2021} because it is clearly observed in the $^{30}$Si($^3$He,t)$^{30}$P study of Ref.~\cite{Ramstein1981} as one component of a doublet ($5799$~keV and $5808$~keV). The triton angular distribution in Fig.~10 of Ref.~\cite{Ramstein1981} is consistent with a transferred angular momentum of $L$ $=$ $4$ and, hence, restricts the spin-parity to $J^\pi$ $=$ $(3-5)^+$ (see Tab.~\ref{tab:results_levels}). 

Although the $E_x$ $=$ $5788$~keV level is not observed in the $^{29}$Si($^3$He,d)$^{30}$P experiment of Ref.~\cite{Dykoski1976}, we can estimate an upper limit for the proton spectroscopic factor by inspecting their Fig.~2, depicting the deuteron spectrum at a laboratory angle of $23^\circ$. Assuming a $3^+$ assignment (Tab.~\ref{tab:results_levels}), we find a value of $C^2S$ $\le$ $0.003$ for a proton orbital angular momentum transfer of $\ell_p$ $=$ $2$, yielding a proton partial width estimate of $\Gamma_p^{195}$ $\le$ $1.5 \times 10^{-8}$~eV (Tab.~\ref{tab:results_resonances}). Smaller values are found for $J^\pi$ assignments of $4^+$ or $5^+$. We will estimate the upper limit contribution of this level to the total reaction rate by assuming a Porter-Thomas probability distribution to sample the dimensionless reduced proton width, and truncating the distribution at a value corresponding to the upper limit of the proton partial width given above. For the mean dimensionless reduced proton width, we adopted a value of $\langle \theta_p^2 \rangle$ $=$ $0.001$, with  an  estimated uncertainty of a factor of $5$ (see Fig.~4 in Ref.~\cite{Pogrebnyak2013}). More information on the treatment of upper limits in Monte Carlo reaction rate sampling is provided in Refs.~\cite{LONGLAND2010,Pogrebnyak2013}.

\subsubsection{$E_r^{cm}$ $=$ $122$~keV, $E_x$ $=$ $5715$~keV [$J^\pi$ $=$ $(5,7)^+$]}\label{sec:5715}
The $E_x$ $=$ $5715$~keV level corresponds to a resonance at $E_r^{cm}$ $=$ $122$~keV. It represents a component of a doublet ($5701$~keV and $5715$~keV), which was not resolved in earlier work \cite{vanGasteren1974,Hafner1974,Boerma1975}, but was resolved in the $^{30}$Si($^3$He,t)$^{30}$P study of Ref.~\cite{Ramstein1981}. This level is also very weakly populated in the $^{29}$Si($^3$He,d)$^{30}$P reaction \cite{Dykoski1976}. The excitation energy recommended by Endt \cite{Endt1998} is a weighted average of those measured by Refs.~\cite{Boerma1975,Dykoski1976,Ramstein1981}. However, Boerma {\it et al.} \cite{Boerma1975} reported a $1^+$ assignment for the level, implying that the peak they observed represents mainly the other component of the doublet ($5701$~keV; see Tab.~\ref{tab:results_levels}). It is not excluded either that the level reported by Dykoski {\it et al.} \cite{Dykoski1976} at ``5711 keV'' corresponds in reality to the lower state ($5701$~keV) of the doublet (see Sec.~\ref{sec:5701}.). Nevertheless, we will adopt here the recommended value of $5715 \pm 4$~keV from ENSDF \cite{ensdf2021}, which was found from the weighted average of the results reported in the ($^3$He,t) \cite{Ramstein1981} and ($^3$He,d) \cite{Dykoski1976} studies (Tabs.~\ref{tab:results_levels} and \ref{tab:results_resonances}). 

Ramstein {\it et al.} \cite{Ramstein1981} reported a transferred angular momentum of $L$ $=$ $6$ in the ($^3$He,t) reaction, implying a spin-parity assignment of $J^\pi$ $=$ $(5,7)^+$. These translate to orbital angular momentum transfers of $\ell_p$ $=$ $4$ ($5^+$) or $\ell_p$ $=$ $6$ ($7^+$) in the $^{29}$Si($^3$He,d)$^{30}$P reaction \cite{Dykoski1976}. As already mentioned, this level is only weakly populated in Ref.~\cite{Dykoski1976}, and they reported no information other than the excitation energy. From the weak peak displayed in their Fig.~2, we can estimate a spectroscopic factor of $C^2S$ $=$ $0.002$ (for $\ell_p$ $=$ $4$) corresponding to a proton partial width of $\Gamma_p$ $\approx$ $3.0 \times 10^{-16}$~eV. Since a $\ell_p$ $=$ $6$ transfer ($5^+$) would yield an even smaller value, we will treat our estimate as an upper limit, $\Gamma_p^{122}$ $\le$ $3.0 \times 10^{-16}$~eV (Tab.~\ref{tab:results_resonances}). The upper limit contribution to the total reaction rate is again obtained by sampling from a Porter-Thomas probability distribution that is truncated at the estimated upper limit of the proton partial width given above.

\subsubsection{$E_r^{cm}$ $=$ $108$~keV, $E_x$ $=$ $5701$~keV [$J^\pi$ $=$ $1^+$]}\label{sec:5701}
The $E_x$ $=$ $5701$~keV level corresponds to an $s$-wave resonance at $E_r^{cm}$ $=$ $108$~keV. For the excitation energy, we adopted the value listed in ENSDF \cite{ensdf2021}, which originates from the measurements reported in Grossmann {\it et al.} \cite{grossmann2000} (see Tab.~\ref{tab:results_levels}). An unambiguous $1^+$ spin-parity assignment is supported by several experiments (see information listed in Refs.~\cite{Endt1998,ensdf2021}). The mean lifetime of this level has also been measured ($\tau_m$ $=$ $16 \pm 4$~fs \cite{Tikkanen1987}), giving a $\gamma$-ray partial width of $\Gamma_{\gamma}$ $=$ $(4.4 \pm 1.1) \times 10^{-2}$~eV.  

As already mentioned in Sec.~\ref{sec:5715}, the very weakly populated level reported in the $^{29}$Si($^3$He,d)$^{30}$P experiment \cite{Dykoski1976} at ``5711 keV'' may correspond
to the $E_x$ $=$ $5701$~keV state. However, in the absence of more information, we can only estimate from Fig.~2 in Ref.~\cite{Dykoski1976} an upper limit of $C^2S$ $\le$ $0.007$ ($\ell_p$ $=$ $0$) for the proton spectroscopic factor, yielding a proton partial width upper limit of $\Gamma_p^{108}$ $\le$ $1.2 \times 10^{-10}$~eV (Tab.~\ref{tab:results_resonances}).

\subsubsection{$E_r^{cm}$ $=$ $3.7$~keV, $E_x$ $=$ $5597$~keV [$J^\pi$ $=$ $4^+$]}\label{sec:5597}
The level at $E_x$ $=$ $5597$~keV is located near the proton threshold. We will adopt the excitation energy listed in ENSDF \cite{ensdf2021}, which originates from the measurement of Ramstein {\it et al.} \cite{Ramstein1981}, instead of the value listed in Endt \cite{Endt1998}, which was mainly adopted from Ref.~\cite{Boerma1975}. The peak near $5594$~keV observed in the latter work sits on top of a large $^{16}$O contamination and may also include a contribution from the nearby level at $5576$~keV ($2^+$). 

The unambiguous spin-parity assignment of the $E_x$ $=$ $5597$~keV state, $J^\pi$ $=$ $4^+$, is based on the observation of a $L$ $=$ $4$ transfer in the $^{30}$Si($^3$He,t)$^{30}$P study of Ref.~\cite{Ramstein1981}. This implies a $\ell_p$ $=$ $4$ transfer in the $^{29}$Si($^3$He,d)$^{30}$P reaction. This state has not been observed by Ref.~\cite{Dykoski1976}, and, therefore, we can only estimate from their Fig.~2 an upper limit for the proton spectroscopic factor of $C^2S$ $\le$ $0.002$.    

Our adopted excitation energy corresponds to a resonance energy of $E_r^{cm}$ $=$ $3.7 \pm 5.0$~keV. At the mean resonance energy value ($3.7$~keV), using the spectroscopic factor estimate, the proton width would amount to only $\Gamma_p^{3.7}$ $\le$ $8.5 \times 10^{-97}$~eV. Therefore, the contribution of this level is irrelevant and it will be disregarded for the calculation of the total reaction rates. 

\subsection{Total rates}\label{sec:total}
The total $^{29}$Si(p,$\gamma$)$^{30}$P thermonuclear reaction rates, based on the Monte Carlo sampling of all experimental nuclear input quantities, are listed in Tab.~\ref{tab:rates} versus stellar temperature. The ``low'', ``median'', and ``high'' rates correspond to the 16, 50, and 84 percentiles, respectively, of the total rate probability density at each temperature. The last column displays the factor uncertainty, $f.u.$, of the total rate. At low ($T$ $\le$ $0.025$~GK) and high ($T$ $\ge$ $0.13$~GK) temperatures, the rate uncertainty is about $50$\% and $\le 20$\%, respectively. Between $0.03$~GK and $0.090$~GK, the uncertainty rises up to an order of magnitude.

%
\begin{table}[ht!] 
\begin{threeparttable}
\caption{Total thermonuclear reaction rates for $^{29}$Si(p,$\gamma$)$^{30}$P \tnote{a}}
\setlength{\tabcolsep}{8pt}
\center
\begin{tabular}{ccccc}
\toprule
T (GK) & Low & Median &   High  &   f.u.  \\
\colrule 
0.010 & 1.43$\times$10$^{-39}$ & 2.10$\times$10$^{-39}$ &
      3.07$\times$10$^{-39}$ & 1.469 \\ 
0.011 & 4.69$\times$10$^{-38}$ & 6.88$\times$10$^{-38}$ &
      9.96$\times$10$^{-38}$ & 1.467 \\ 
0.012 & 1.02$\times$10$^{-36}$ & 1.49$\times$10$^{-36}$ &
      2.20$\times$10$^{-36}$ & 1.475 \\ 
0.013 & 1.61$\times$10$^{-35}$ & 2.36$\times$10$^{-35}$ &
      3.45$\times$10$^{-35}$ & 1.467 \\ 
0.014 & 1.92$\times$10$^{-34}$ & 2.82$\times$10$^{-34}$ &
      4.13$\times$10$^{-34}$ & 1.471 \\ 
0.015 & 1.85$\times$10$^{-33}$ & 2.70$\times$10$^{-33}$ &
      3.97$\times$10$^{-33}$ & 1.467 \\ 
0.016 & 1.46$\times$10$^{-32}$ & 2.14$\times$10$^{-32}$ &
      3.13$\times$10$^{-32}$ & 1.467 \\ 
0.018 & 5.69$\times$10$^{-31}$ & 8.29$\times$10$^{-31}$ &
      1.22$\times$10$^{-30}$ & 1.466 \\ 
0.020 & 1.38$\times$10$^{-29}$ & 2.00$\times$10$^{-29}$ &
      2.92$\times$10$^{-29}$ & 1.458 \\ 
0.025 & 1.46$\times$10$^{-26}$ & 4.93$\times$10$^{-26}$ &
      2.38$\times$10$^{-25}$ & 3.399 \\ 
0.030 & 1.01$\times$10$^{-23}$ & 1.24$\times$10$^{-22}$ &
      7.32$\times$10$^{-22}$ & 7.074 \\ 
0.040 & 1.96$\times$10$^{-19}$ & 2.72$\times$10$^{-18}$ &
      1.62$\times$10$^{-17}$ & 9.509 \\ 
0.050 & 7.36$\times$10$^{-17}$ & 1.03$\times$10$^{-15}$ &
      6.12$\times$10$^{-15}$ & 9.932 \\ 
0.060 & 3.65$\times$10$^{-15}$ & 5.10$\times$10$^{-14}$ &
      3.04$\times$10$^{-13}$ & 9.797 \\ 
0.070 & 5.86$\times$10$^{-14}$ & 8.01$\times$10$^{-13}$ &
      4.76$\times$10$^{-12}$ & 8.810 \\ 
0.080 & 5.19$\times$10$^{-13}$ & 6.22$\times$10$^{-12}$ &
      3.66$\times$10$^{-11}$ & 7.053 \\ 
0.090 & 4.34$\times$10$^{-12}$ & 3.15$\times$10$^{-11}$ &
      1.77$\times$10$^{-10}$ & 5.041 \\ 
0.100 & 4.55$\times$10$^{-11}$ & 1.41$\times$10$^{-10}$ &
      6.39$\times$10$^{-10}$ & 3.180 \\ 
0.110 & 5.45$\times$10$^{-10}$ & 8.54$\times$10$^{-10}$ &
      2.18$\times$10$^{-09}$ & 1.907 \\ 
0.120 & 5.79$\times$10$^{-09}$ & 7.08$\times$10$^{-09}$ &
      1.00$\times$10$^{-08}$ & 1.326 \\ 
0.130 & 4.68$\times$10$^{-08}$ & 5.40$\times$10$^{-08}$ &
      6.33$\times$10$^{-08}$ & 1.164 \\ 
0.140 & 2.92$\times$10$^{-07}$ & 3.33$\times$10$^{-07}$ &
      3.80$\times$10$^{-07}$ & 1.142 \\ 
0.150 & 1.46$\times$10$^{-06}$ & 1.66$\times$10$^{-06}$ &
      1.89$\times$10$^{-06}$ & 1.140 \\ 
0.160 & 6.00$\times$10$^{-06}$ & 6.81$\times$10$^{-06}$ &
      7.75$\times$10$^{-06}$ & 1.139 \\ 
0.180 & 6.37$\times$10$^{-05}$ & 7.21$\times$10$^{-05}$ &
      8.20$\times$10$^{-05}$ & 1.136 \\ 
0.200 & 4.23$\times$10$^{-04}$ & 4.77$\times$10$^{-04}$ &
      5.40$\times$10$^{-04}$ & 1.132 \\ 
0.250 & 1.29$\times$10$^{-02}$ & 1.44$\times$10$^{-02}$ &
      1.61$\times$10$^{-02}$ & 1.119 \\ 
0.300 & 1.29$\times$10$^{-01}$ & 1.43$\times$10$^{-01}$ &
      1.58$\times$10$^{-01}$ & 1.106 \\ 
0.350 & 6.82$\times$10$^{-01}$ & 7.47$\times$10$^{-01}$ &
      8.20$\times$10$^{-01}$ & 1.097 \\ 
0.400 & 2.39$\times$10$^{+00}$ & 2.61$\times$10$^{+00}$ &
      2.85$\times$10$^{+00}$ & 1.092 \\ 
0.450 & 6.35$\times$10$^{+00}$ & 6.92$\times$10$^{+00}$ &
      7.54$\times$10$^{+00}$ & 1.090 \\ 
0.500 & 1.38$\times$10$^{+01}$ & 1.51$\times$10$^{+01}$ &
      1.64$\times$10$^{+01}$ & 1.089 \\ 
0.600 & 4.39$\times$10$^{+01}$ & 4.79$\times$10$^{+01}$ &
      5.22$\times$10$^{+01}$ & 1.090 \\ 
0.700 & 9.86$\times$10$^{+01}$ & 1.08$\times$10$^{+02}$ &
      1.18$\times$10$^{+02}$ & 1.092 \\ 
0.800 & 1.78$\times$10$^{+02}$ & 1.95$\times$10$^{+02}$ &
      2.13$\times$10$^{+02}$ & 1.093 \\ 
0.900 & 2.80$\times$10$^{+02}$ & 3.07$\times$10$^{+02}$ &
      3.35$\times$10$^{+02}$ & 1.093 \\ 
1.000 & 3.99$\times$10$^{+02}$ & 4.37$\times$10$^{+02}$ &
      4.78$\times$10$^{+02}$ & 1.093 \\ 
1.250 & 7.39$\times$10$^{+02}$ & 8.07$\times$10$^{+02}$ &
      8.81$\times$10$^{+02}$ & 1.092 \\ 
1.500 & 1.10$\times$10$^{+03}$ & 1.19$\times$10$^{+03}$ &
      1.30$\times$10$^{+03}$ & 1.090 \\ 
1.750 & 1.44$\times$10$^{+03}$ & 1.57$\times$10$^{+03}$ &
      1.71$\times$10$^{+03}$ & 1.088 \\ 
2.000 & 1.77$\times$10$^{+03}$ & 1.92$\times$10$^{+03}$ &
      2.09$\times$10$^{+03}$ & 1.085 \\ 
2.500 & 2.42$\times$10$^{+03}$ & 2.61$\times$10$^{+03}$ &
      2.82$\times$10$^{+03}$ & 1.079 \\ 
3.000 & 3.10$\times$10$^{+03}$ & 3.31$\times$10$^{+03}$ &
      3.55$\times$10$^{+03}$ & 1.072 \\ 
3.500 & 3.83$\times$10$^{+03}$ & 4.07$\times$10$^{+03}$ &
      4.34$\times$10$^{+03}$ & 1.066 \\ 
4.000 & 4.60$\times$10$^{+03}$ & 4.88$\times$10$^{+03}$ &
      5.18$\times$10$^{+03}$ & 1.062 \\ 
5.000 & 6.23$\times$10$^{+03}$ & 6.57$\times$10$^{+03}$ &
      6.94$\times$10$^{+03}$ & 1.057 \\ 
\botrule \\
\end{tabular}
\begin{tablenotes}
\item[a] {\footnotesize In units of cm$^3$mol$^{-1}$s$^{-1}$. Columns 2, 3, and 4 list the 16th, 50th, and 84th percentiles of the total rate probability density at given temperatures; $f.u.$ is the factor uncertainty, based on Monte Carlo sampling, of the total reaction rate. The total number of samples at each temperature was 20,000.}
\end{tablenotes}
\label{tab:rates}
\end{threeparttable}
\end{table}    

\begin{figure}[!ht]
\includegraphics[width=1.0\columnwidth]{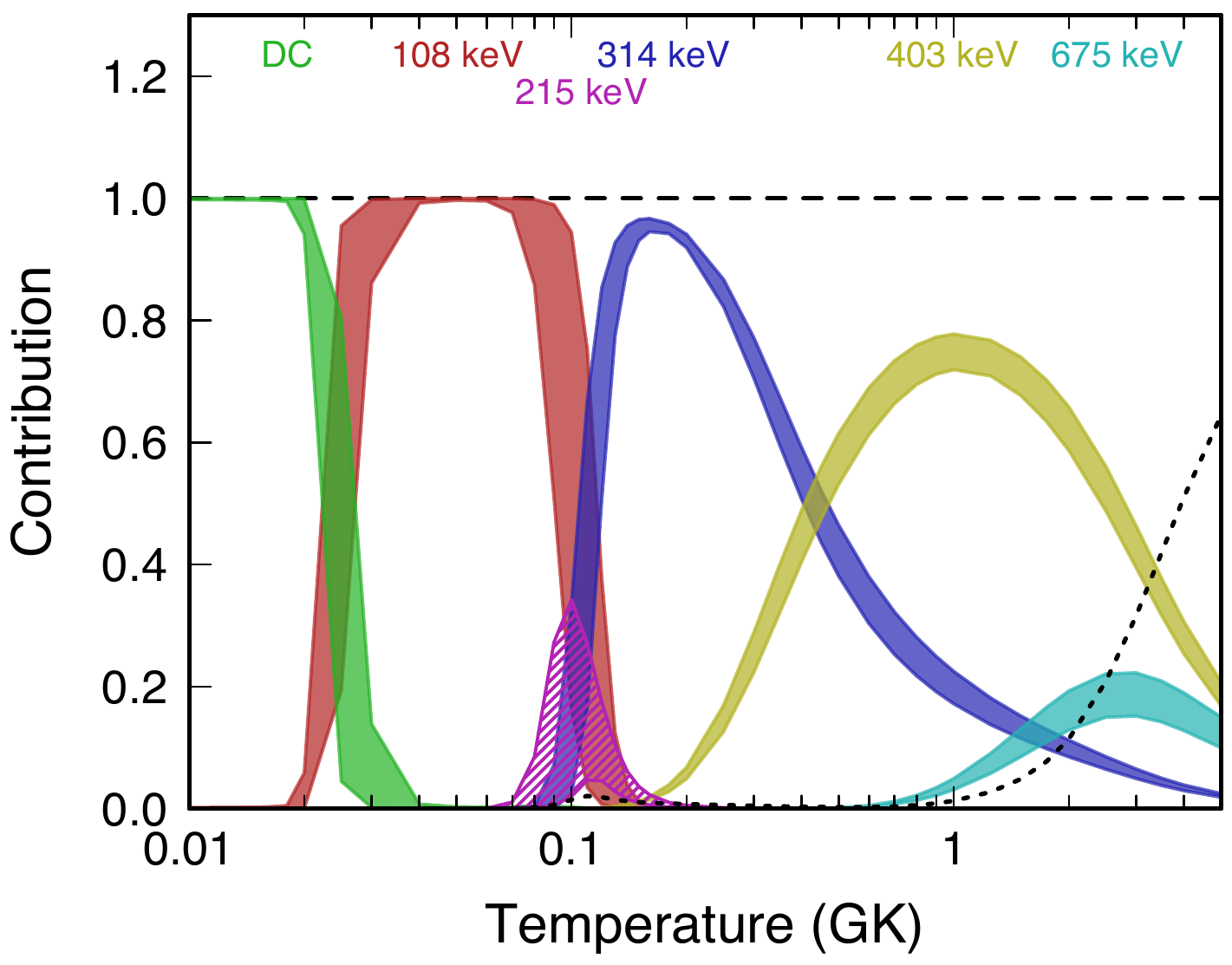}
\caption{\label{fig:contributions} 
The fractional contributions to the total $^{29}$Si(p,$\gamma$)$^{30}$P reaction rate. Resonances are labeled by their center-of-mass energies and the label ``DC'' refers to the direct proton capture process (Sec.~\ref{sec:dc}). The contribution ranges are shown as colored bands, with the band thickness representing the uncertainty of the contribution. The dotted black line corresponds to the contribution of resonances with energies larger than $700$~keV. 
}
\end{figure}

The fractional rate contributions (i.e., relative to the total rate) are displayed in Fig.~\ref{fig:contributions}. The energy labels refer to center-of-mass resonance energies, while ``DC'' represents the direct capture contribution. The latter process dominates the rates below $0.020$~GK. Between $0.025$~GK and $0.1$~GK, the unobserved $E_r^{cm}$ $=$ $108$~keV resonance ($E_x$ $=$ $5701$~keV; Sec.~\ref{sec:5701}) dominates the total rates. The resonances at $E_r^{cm}$ $=$ $314$~keV ($E_x$ $=$ $5908$~keV) and $E_r^{cm}$ $=$ $403$~keV ($E_x$ $=$ $5996$~keV), both measured in the present work (Sec.~\ref{sec:results}), are the major contributors to the total rate in the $0.13$ $-$ $0.4$~GK and $0.4$ $-$ $2.0$~GK temperature ranges, respectively. At higher temperatures, resonances with energies above $500$~keV dominate the total rate. It can also be seen in Fig.~\ref{fig:contributions} that the unobserved $E_r^{cm}$ $=$ $215$~keV resonance ($E_x$ $=$ $5808$~keV), for which we measured an upper limit for the resonance strength (Secs.~\ref{sec:215} and \ref{sec:5808}), contributes up to $\approx 35$\% to the total rate in the narrow temperature window of $T$ $=$ $0.08$ $-$ $0.12$~GK. 

The rate contributions of three low-energy resonances (Tab.~\ref{tab:results_resonances}) are negligible. The $E_r^{cm}$ $=$ $122$~keV resonance has a high spin ($J$ $=$ $5$), implying a $g$-wave resonance and a small proton partial width. The $E_r^{cm}$ $=$ $195$-keV resonance is too close to the much stronger $215$-keV resonance for which we estimated an actual resonance strength value rather than an upper limit (Sec.~\ref{sec:5808} and Tab.~\ref{tab:results_resonances}). And the $E_r^{cm}$ $=$ $303$-keV resonance, which we first observed in the present work (Sec.~\ref{sec:303}), has a much smaller strength than that of the nearby $314$~keV resonance.   

\begin{figure}[!ht]
\includegraphics[width=1.0\columnwidth]{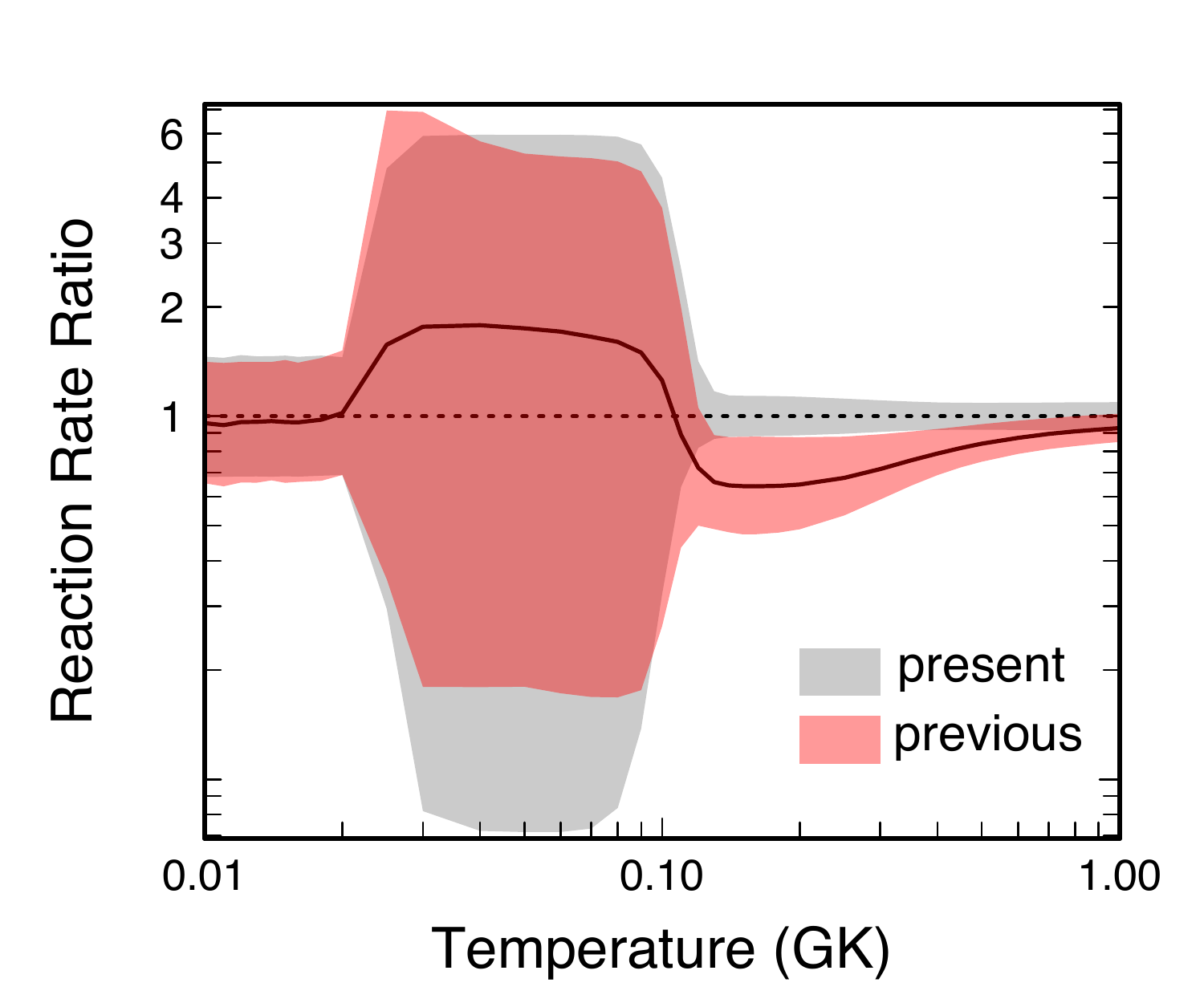}
\caption{\label{fig:ratecomparison} 
Reaction rates for $^{29}$Si(p,$\gamma$)$^{30}$P from the present work (gray) and the 2010 evaluation of Ref.~\cite{ILIADIS2010b} (red), normalized to the present recommended (median) rates. The shaded areas correspond to 68\% coverage probabilities. The black solid line shows the ratio of previous to present recommended rates. Notice that the present recommended $^{29}$Si $+$ $p$ rates at temperatures of $T$ $=$ $0.13$ $-$ $0.4$~GK, which are most important for classical nova nucleosynthesis, are higher than the previous result by up to $50$\%, and have also a much smaller uncertainty. See text.
}
\end{figure}

Figure~\ref{fig:ratecomparison} compares the present rates (gray) with the results of the 2010 evaluation of Ref.~\cite{ILIADIS2010b} (red). The boundaries of the shaded regions correspond to the 16 and 84 percentiles of the total rate probability density distributions. All rates shown are normalized to the present recommended rate. The solid black line corresponds to the ratio of the two recommended rates. Two aspects are noteworthy. First, at temperatures of $T$ $=$ $0.13$ $-$ $0.4$~GK, which are most important for classical nova nucleosynthesis, the present recommended rate is higher than the previous result by up to $50$\%, and has also a much smaller uncertainty. At these temperatures, the total rate is dominated by the $E_r^{cm}$ $=$ $314$~keV resonance (see Fig.~\ref{fig:contributions}), which we measured in the present work (Sec.~\ref{sec:314}) and for which we obtained a much improved strength (Tab.~\ref{tab:results_resonances}). Second, at lower temperatures, $T$ $=$ $0.03$ $-$ $0.09$~GK, where the reaction rate is dominated by the $E_r^{cm}$ $=$ $108$~keV resonance (see Fig.~\ref{fig:contributions}), the new rate has a {\it larger} uncertainty than the 2010 result. One reason is that our estimated proton partial width (Tab.~\ref{tab:results_resonances}) exceeds the 2010 value by a factor of $2$. Furthermore, to estimate the upper limit contributions of resonances we adopted $\langle \theta_p^2 \rangle$ $=$ $0.001$ for the mean value of the dimensionless proton reduced width (Sec.~\ref{sec:5788}), compared to a value of $0.00045$ used in the 2010 evaluation. Our value is based on Fig.~4 of Ref.~\cite{Pogrebnyak2013}, which was published after the 2010 evaluation. Also, we used the most recent version of \texttt{RatesMC} (v. 2.11.0), which allowed us to input an uncertainty for this quantity (i.e., a factor of $5$).

The rates presented in Tab.~\ref{tab:rates} were estimated without taking any correlations in the resonance strengths into account. However, these quantities are correlated because we normalized all strengths to the value of the
$E_r^{cm}$ $=$ $403$~keV resonance recommended by Sargood \cite{SARGOOD198261} (see Sec.~\ref{sec:ratereshigh}). Tests have shown that, when correlations are fully taken into account using the method presented in Ref.~\cite{longland2017}, the $^{29}$Si $+$ $p$ recommended rate, e.g., at $200$~MK, remains unchanged, while the rate uncertainty increases slightly from 13.2\% (Tab.~\ref{tab:rates}) to 14.4\%. This small change is negligible.

\subsection{Numerical tests}\label{sec:tests}
The reaction rates presented here have been obtained by Monte Carlo sampling of all experimental uncertainties. Therefore, they represent our best estimate, based on all {\it known} statistical and systematic effects \cite{LONGLAND2010}. In particular, if only an upper limit was obtained for a proton partial width, we are drawing samples for the corresponding reduced width (or spectroscopic factor) from a Porter-Thomas distribution \cite{PorterThomas} (i.e., a chi-squared distribution with one degree of freedom), as first suggested by Ref.~\cite{Thompson99}. This assumption is informed by the nuclear statistical model, and implies a higher probability of sampling the smaller the value of the reduced width. For details see Refs.~\cite{LONGLAND2010,Pogrebnyak2013}. However, as pointed out in Ref.~\cite{LONGLAND2010}, the Monte Carlo-based reaction rate estimate does not include {\it unknown} systematic effects. Suppose that the reduced widths of levels in a local region are not statistically distributed, e.g., that a given level carries for some reason a particularly large fraction of the single-particle strength. In this case, we would underestimate its rate contribution by sampling from a Porter-Thomas distribution of reduced widths. 

We will now address a number of related questions. How robust are our recommended reaction rates (Tab.~\ref{tab:rates}) if the levels for which we could only estimate upper limit contributions would have spectroscopic factors just below their experimental upper limits? Is it worthwhile to perform future experimental searches for unobserved resonances? Will the reaction rates change in the temperature region important for classical novae ($T$ $\ge$ $0.13$~GK) if we take such unknown systematic effects into account? 

To answer these questions, we performed a series of numerical tests. Our findings are summarized below:

(i) Recall that for the $E_r^{cm}$ $=$ $215$~keV resonance, we could only obtain an experimental upper limit for the resonance strength (Sec.~\ref{sec:215}). However, with the additional information provided by the proton-transfer measurement \cite{Dykoski1976}, we estimated the proton partial width, $\Gamma_p^{215}$ $\approx$ $1.0 \times 10^{-7}$~eV, with a factor of $2$ uncertainty (Sec.~\ref{sec:5808}). If a future $^{29}$Si $+$ $p$ measurement would detect this resonance consistent with our estimated strength, but determine a much smaller resonance strength uncertainty (say, $\pm 25$\%), the total rates or their uncertainties near $T$ $\approx$ $0.1$~GK would change by less than $15$\%. 

(ii) For the $E_r^{cm}$ $=$ $122$~keV and $195$~keV resonances, either setting their proton partial widths right at their respective upper limit or disregarding them entirely, changes the total rates by only a few percent. Therefore, future searches for these resonances are not urgent for improving the total rate.

\begin{figure}[!ht]
\includegraphics[width=1.0\columnwidth]{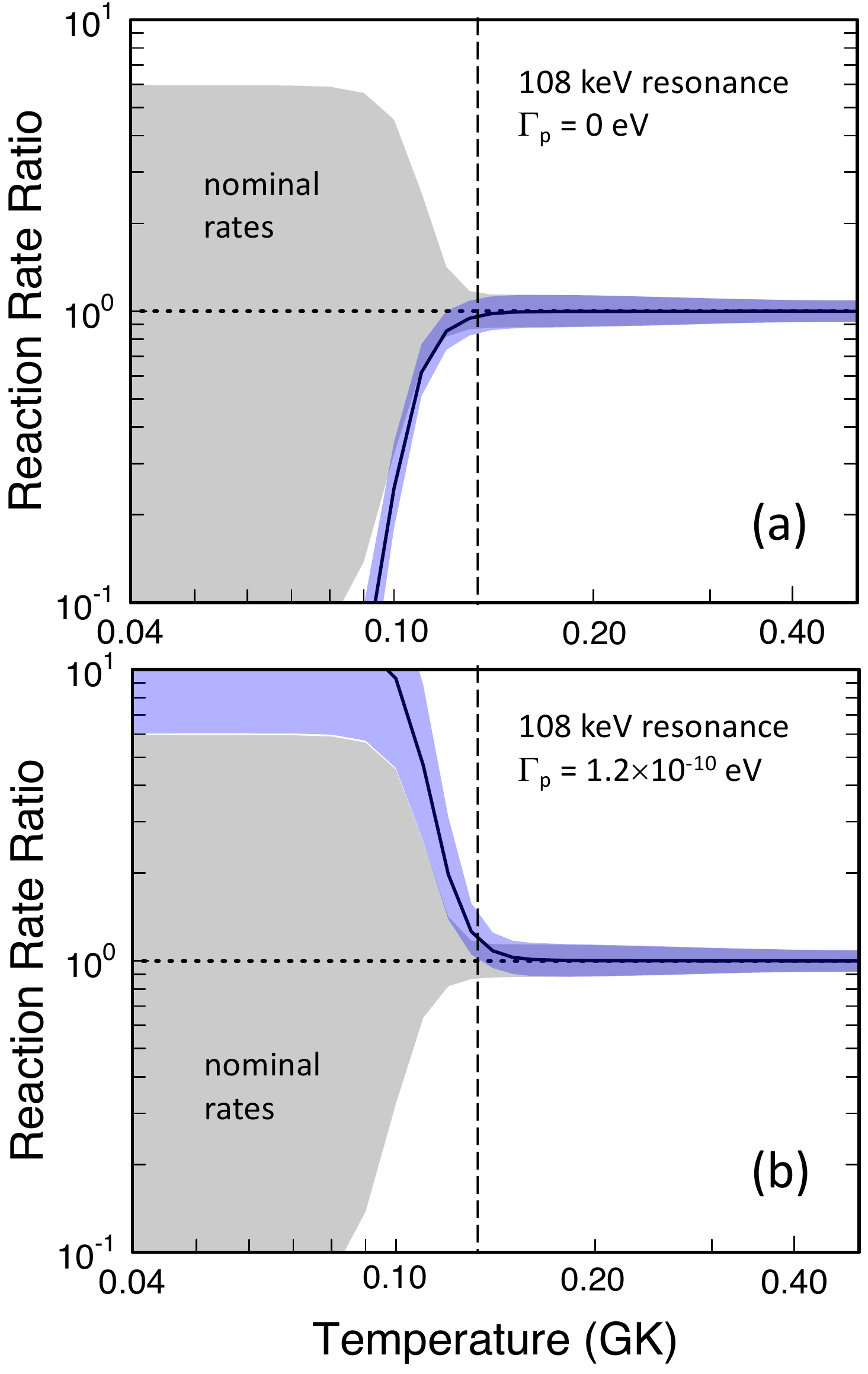}
\caption{\label{fig:108comp} 
Test results to explore the impact of the unobserved $E_r^{cm}$ $=$ $108$~keV resonance on the total rates at classical nova temperatures. The gray shaded region represents the nominal rates listed in Tab.~\ref{tab:rates} and is the same as the gray shaded area in Fig.~\ref{fig:ratecomparison}. The blue shaded regions are the result of repeating the rate calculation assuming the $108$-keV resonance (a) does not exist, or (b) has a maximum strength allowed by our estimated upper limit (Tab.~\ref{tab:results_resonances}). The vertical dashed line corresponds to a temperature of $T$ $=$ $0.13$~GK. Nova nucleosynthesis usually takes place at temperatures to the right of the dashed line. At lower temperatures, the $108$-keV resonance will significantly impact the total rates.  
}
\end{figure}

(iii) As mentioned above, the $E_r^{cm}$ $=$ $108$~keV resonance strongly impacts the rates at lower temperatures. Figure~\ref{fig:108comp} illustrates how the total rate (blue shaded region) is affected if we assume that the resonance does not exist ($\Gamma_p^{108}$ $=$ $0$~eV; top panel), or has the maximum strength consistent with our estimated upper limit ($\Gamma_p^{108}$ $\approx$ $1.2 \times 10^{-10}$~eV; bottom panel; see Tab.~\ref{tab:results_resonances} and Sec.~\ref{sec:5701}). At temperatures of $T$ $\ge$ $0.13$~GK the rates change by less than 25\% compared to our nominal rate (gray shaded region). At decreasing temperatures, the differences with respect to our nominal rates become quickly larger (e.g., at $0.12$~GK they amount to a factor of $2$). A future proton transfer study to measure an improved value for the spectroscopic factor of this $s$-wave resonance would be important if more precise $^{29}$Si $+$ $p$ reaction rates are desired at temperatures below $0.13$~GK.

%

\section{Comparison of results to Ref.~\cite{Lotay2020}}\label{sec:lotay}
Lotay et al. \cite{Lotay2020} used the $^{28}$Si($^3$He,p) fusion-evaporation reaction to measure the $\gamma$-ray decay of populated $^{30}$P levels using the Gammasphere detector array. They presented experimental $^{30}$P excitation energies and spin-parities, and estimated $^{29}$Si $+$ $p$ strengths (or upper limits) of low-energy resonances indirectly.

We did not include their results in our experimental averages (Tabs.~\ref{tab:results_levels} $-$ \ref{tab:results_resonances}), which are based on the data evaluations of Refs.~\cite{Endt1998,ensdf2021} and results of the present work, because we were unable to discern from the information given in Tab.~I of Ref.~\cite{Lotay2020} the extent to which their spin-parities are based on shell-model calculations or previous studies.
Future evaluations will consider all available experimental information and recommend the most up-to-date nuclear structure information. Here, we will briefly comment on those results in Ref.~\cite{Lotay2020} that are related to the $^{29}$Si $+$ $p$ reaction rate.

The most important prediction in Ref.~\cite{Lotay2020} pertains to the strength of the $E_r^{cm}$ $=$ $314$~keV resonance (Sec.~\ref{sec:314}), which was unobserved prior to the present work. To calculate the resonance strength, Ref.~\cite{Lotay2020} used the average of the spectroscopic factor values reported in Refs.~\cite{Dykoski1976,Hertzog1974} and found, from Eq.~(3), a strength of $\omega\gamma^{Lotay}$ $\approx$ $(2J+1)\Gamma_p/4$ $=$ $0.025$~eV. Such indirect estimates carry relatively large uncertainties\footnote{The strength reported by Ref.~\cite{Lotay2020} is $\omega\gamma^{Lotay}$ $=$ $0.025 \pm 0.004$~eV. Their small uncertainty ($16$\%) was obtained from the $15$\% uncertainty ascribed to the extraction of the absolute cross section in Ref.~\cite{Dykoski1976}. However, besides the absolute cross section scale, the Distorted Wave Born Approximation (DWBA) formalism used to analyze the stripping cross sections introduces another significant source of uncertainty. A systematic comparison of spectroscopic factors in the $A$ $=$ $21$ $-$ $44$ region by Endt \cite{Endt1977} and in the sd-shell by Ref.~\cite{Thompson99} resulted in a $25$\% and $40$\% experimental uncertainty, respectively, for individual measurements of {\it strong transitions}.} and, therefore, their result encompasses all directly-measured values, including that of the present work (see also Sec.~\ref{sec:314} and Tab.~\ref{tab:results_resonances}, and references therein).

For the $E_r^{cm}$ $=$ $303$~keV resonance, Ref.~\cite{Lotay2020} estimated an upper strength limit of $\omega\gamma^{Lotay}$ $\le$ $4.7 \times 10^{-5}$~eV by guessing a spectroscopic factor of $C^2S$ $\le$ $0.01$. Their upper limit estimate is significantly smaller than our directly measured strength, $\omega\gamma$ $=$ $(8.8 \pm 1.5) \times 10^{-5}$~eV (Tab.~\ref{tab:results_resonances}). Furthermore, Ref.~\cite{Lotay2020} reported an unambiguous spin-parity of $3^+$ for the corresponding level at $E_x$ $=$ $5897$~keV, whereas we find $J^\pi$ $=$ $(1,2)^+$, based on the decays observed in our spectra (Fig.~\ref{fig:spec} and Tab.~\ref{tab:results_levels}). In particular, we clearly observe the primary transition $R \rightarrow 677$~keV in both singles and coincidence spectra, as can be seen in Fig.~\ref{fig:spec}. Since the $E_x$ $=$ $677$~keV final state has a spin-parity of $0^+$, a $3^+$ assignment to the decaying $E_x$ $=$ $5897$~keV state would imply an unlikely $M3$ transition.

Finally, for the $E_r^{cm}$ $=$ $215$~keV resonance, Ref.~\cite{Lotay2020} reports an unambiguous spin-parity of $2^+$ for the corresponding level at $E_x$ $=$ $5808$~keV, which is inconsistent with the $(3,5)^+$ assignment listed in the evaluations of Refs.~\cite{Endt1998,ensdf2021}. We assigned a value of $J^\pi$ $=$ $3^+$ to this level, based on the discussion in Sec.~\ref{sec:215}. From the measured spectroscopic factor \cite{Dykoski1976}, Ref.~\cite{Lotay2020} estimated a resonance strength of $\omega\gamma^{Lotay}$ $=$ $(4.7 \pm 0.7) \times 10^{-7}$~eV, which is higher than our directly measured upper limit of $\omega\gamma$ $\le$ $3.3 \times 10^{-7}$~eV (Tab.~\ref{tab:results_resonances}). One likely reason for the discrepancy is that Ref.~\cite{Lotay2020} has underestimated the uncertainty in the spectroscopic factor ($17$\%). As we already mentioned, experimental spectroscopic factors for strong transitions carry a combined uncertainty (cross section scale and reaction formalism) of $\approx 25$\% $-$ $40$\%, and a likely higher uncertainty for weak transitions such as in this case. We also point out that measured stripping reaction angular distributions are sensitive to the orbital angular momentum transfer, but not to a specific value of the spin, $J$. For this reason, such studies, including Ref.~\cite{Dykoski1976}, report their results as $(2J + 1) C^2S$. Since the factor $(2J + 1)$ appears also in the definition of $\omega\gamma$, the value of the spin, $J$, is irrelevant for the strength estimate of the $E_r^{cm}$ $=$ $215$~keV resonance.

\section{Summary and conclusions}\label{sec:summary}
Accurate knowledge of the $^{29}$Si(p,$\gamma$)$^{30}$P thermonuclear rate is required to interpret silicon isotopic ratios measured in presolar stardust grains in primitive meteorites. We measured three resonances, at center-of-mass energies of $E_r^{cm}$ $=$ $303$~keV, $314$~keV, and $403$~keV, and obtained improved values for the resonance energies and strengths, excitation energies, branching ratios, and spin-parities. The first of these resonances had not been measured previously. We also searched for an unobserved resonance at $E_r^{cm}$ $=$ $215$~keV, and obtained an experimental upper limit for the resonance strength. The nuclear structure information near the proton threshold was evaluated and total reaction rates were estimated based on all the available experimental information. 

We found that, at the temperatures of $T$ $=$ $0.13$ $-$ $0.4$~GK most important for classical novae, the present $^{29}$Si $+$ $p$ reaction rates are higher than previous estimates, by up to $50$\%, and also have much smaller uncertainties (reduction by a factor up to $2.5$). In this temperature range, our recommended rates are not expected to change significantly if the partial widths of as yet unobserved resonances are systematically varied within experimentally allowed ranges. For lower temperatures, the $E_r^{cm}$ $=$ $108$~keV $s$-wave resonance could change our recommended rates by orders of magnitude. A future proton-transfer measurement of the spectroscopic factor for this resonance could significantly reduce the rate uncertainties at $T$ $<$ $0.13$~GK. The astrophysical implications of our new $^{29}$Si(p,$\gamma$)$^{30}$P rates have been presented in Ref.~\cite{downen22}.

\begin{acknowledgments}
We would like to thank Andrew Cooper and Sean Hunt for help with the JN and ECR measurements. The comments of Robert Janssens, Richard Longland, and Filip Kondev are highly appreciated. This work is supported by the DOE, Office of Science, Office of Nuclear Physics, under grants DE-FG02-97ER41041 (UNC) and DE-FG02-97ER41033 (TUNL). 
\end{acknowledgments}

\bibliography{paper}

\end{document}